\documentclass[a4paper,11pt]{article}
\pdfoutput=1 % if your are submitting a pdflatex (i.e. if you have
\usepackage{jheppub} % for details on the use of the package, please
\usepackage{amsmath}
\allowdisplaybreaks[4]        
\usepackage{amssymb}
\usepackage{euscript}     
\usepackage{color}         
\usepackage{tensor}        
\usepackage{amsthm}

\usepackage{graphicx,subcaption}
 \usepackage{mathtools}

\usepackage{hyperref}
\hypersetup{colorlinks=true,allcolors=blue}

\usepackage[most]{tcolorbox}
\tcbuselibrary{skins, breakable, theorems}

\usepackage[header,title,page,titletoc]{appendix}  
\usepackage[T1]{fontenc} % if needed
\usepackage[numbers]{natbib}

\newcommand{\arccosh}{\text{arccosh}}

\definecolor{browna}{rgb}{0.76,0.72,0.65}
\definecolor{brownb}{rgb}{0.71,0.69,0.65}
\definecolor{SpringGreen}{rgb}{0.95,0.97,0.95}
\definecolor{OliverGreen}{rgb}{0.09,0.34,0.09}
\definecolor{LeftGreen}{rgb}{0.13,0.54,0.13}

\newtcbtheorem[no counter]{theorem}{Theorem}{
	enhanced,
	sharp corners,
	attach boxed title to top left={
		yshifttext=-1mm
	},
	colback=white,
	colframe=browna,
	fonttitle=\bfseries,
	boxed title style={
		sharp corners,
		size=small,
		colback=browna,
		colframe=browna,
	} 
}{thm}

\newtcbtheorem[no counter]{Example}{Example}{
	enhanced,
	sharp corners,
	attach boxed title to top left={
		yshifttext=-1mm
	},
	colback=white,
	colframe=brownb,
	fonttitle=\bfseries,
	coltitle=white,
	boxed title style={
		sharp corners,
		size=small,
		colback=brownb,
		colframe=brownb,
	} 
}{exm}

\newtcbtheorem{myclaim}{Claim -- }
{    coltitle=OliverGreen,    
	colback=SpringGreen,    colframe=LeftGreen,    detach title,    boxrule=0pt,    leftrule=2pt,    attach title to upper,    sharp corners,    left=1mm,}
{claim}

\definecolor{browna}{rgb}{0.76,0.72,0.65}

\subheader{\today}

\title{Non-perturbative Discrete Spectrum of Interior Length and Timeshift in Two-sided Black Hole} 

\author{Masamichi Miyaji}

\affiliation{Yukawa Institute for Theoretical Physics, Kyoto University, Kyoto, 606-8267, Japan}
\emailAdd{masamichi.miyaji@gmail.com}

\abstract{We study the spectrum of the interior length and the horizon timeshift of a two-sided black hole by constructing non-perturbative length and timeshift operators in Jackiew-Teitelboim gravity. We first construct projection operators onto the fixed length or fixed horizon timeshift subspaces using the replica trick. We calculate the densities of state for the length and the timeshift, which are found to be finite and terminate at values of order $e^{S_0}$. This finiteness implies the discreteness in the spectrum of these quantities, and significant modifications in length and timeshift spectrum at order $e^{S_0}$. We then construct the non-perturbative length and timeshift operators, and apply them to study the time evolution of the two-sided black hole. We find that at early time, the probability distribution of the interior length and the timeshift are sharply peaked at the classical values, while after the Heisenberg time, the distribution is completely uniform over all possible values of the length and the timeshift, indicating maximal uncertainty. In particular, the probability of having the negative timeshift states, which corresponds to the white hole probability, is $O(1)$ after the Heisenberg time.}

\newcommand{\imineq}[2]{\vcenter{\hbox{\includegraphics[height=#2ex]{#1}}}}

\numberwithin{equation}{section}

\begin{document}
%%%%%%%%%%%%%%%%%%%%%%%%%%%%%%%%%%%%%%%%%%%%%%%%%%%%
\begin{flushright}
YITP-24-139
\\
\end{flushright}
%%%%%%%%%%%%%%%%%%%%%%%%%%%%%%%%%%%%%%%%%%%%%%%%%%%%
	\maketitle
	\flushbottom
	
\section{Introduction}

Inclusion of Euclidean wormhole \cite{Coleman:1988cy, Giddings:1987cg} into the gravitational path-integral turned out to be powerful approach to study generic properties of quantum gravity, maintaining both intuitive spacetime geometry as well as the finiteness of the quantum gravity Hilbert space dimension \cite{Saad:2018bqo, Saad:2019lba, Penington:2019npb, Almheiri:2019psf, Almheiri:2019hni, Penington:2019kki, Almheiri:2019qdq, Saad:2019pqd, Okuyama:2020ncd, Blommaert:2022lbh, Saad:2022kfe, Okuyama:2023pio}. Combining with the connection to the quantum entanglement \cite{Ryu:2006bv, Hubeny:2007xt, Faulkner:2013ana, Hartman:2013qma, Lewkowycz:2013nqa, Jafferis:2015del}, it is possible to extract bulk degrees of freedom from subregion of the AdS boundary \cite{Czech:2012bh, Pastawski:2015qua, Harlow:2016vwg, Dong:2016eik, Akers:2022qdl}
or the interior information from the Hawking radiation.

On the other hand, the existence of Euclidean wormholes implies that the gravity quantum states on fixed time slice geometry are not necessarily orthogonal with each other due to non-factorized geometry, even if they can never be connected via the change of slicing \cite{Saad:2019pqd, Iliesiu:2021ari, Iliesiu:2024cnh}. This is due to the effect of baby universe emission, which can change the geometry of the timeslice \cite{Coleman:1988cy,Giddings:1987cg,Saad:2019pqd}, or equivalently due to the presence of null states \cite{Coleman:1988cy, Marolf:2020xie, Marolf:2020rpm}. The gravitational path-integral with Euclidean wormholes in AdS is dual to ensemble averaging over boundary theories \cite{Saad:2018bqo, Bousso:2019ykv, Bousso:2020kmy}, and corresponds to the ensemble average over microscopic gravity theories in the bulk \cite{Saad:2021uzi, Blommaert:2021fob, Blommaert:2022ucs}. The ensemble average can indeed reproduce typical behavior of the gravity theories in the ensemble such as the entropy of Hawking radiation \cite{Bousso:2023efc} and the recoverability of the interior \cite{Miyaji:2023wcf} by confirming the deviations are exponentially suppressed. It is possible to reproduce low energy density of states of near extremal black holes \cite{Iliesiu:2020qvm, Heydeman:2020hhw, Iliesiu:2022kny, Boruch:2023trc} as well as the factorization of bulk Hilbert space \cite{Boruch:2024kvv, Balasubramanian:2024yxk}.

The natural way to construct quantum gravity states with fixed timeslice geometry is to use an extension of the Hartle-Hawking prescription \cite{Hartle:1983ai}. We denote the values of geometric quantities as $f$, and corresponding geometric states as $|f\rangle$. Then the ensemble average of the product of overlaps between geometric states is
\begin{equation}
    \mathbb{E}\Big[\langle\text{TFD}|f_1\rangle\langle f_1'|\text{TFD}\rangle\cdots \Big]
    =\sum_{\text{Wormholes}}\imineq{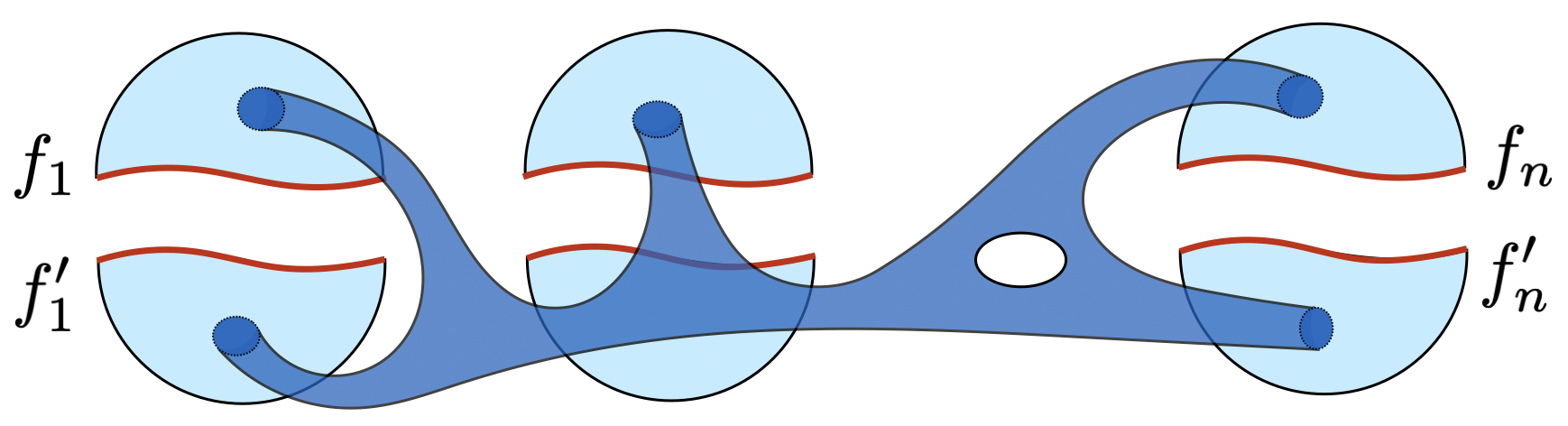}{10}.
\end{equation}
Here the sum is over all possible Euclidean wormhole contributions with fixed boundary geometry. In particular, the Euclidean wormholes do not change the topology of the timeslice. The effect of Euclidean wormholes is that $\langle f|f'\rangle$ is no longer orthogonal even if it was so in the absence of Euclidean wormholes. The immediate consequence is that the notion of the geometry of timeslice becomes ambiguous as an observable since it is no longer an eigenvalue of an Hermitian operator. The same issue appeared in the end-of-the-world brane models of black holes, which have many excitations behind horizon that are highly non-orthogonal with each other \cite{Penington:2019kki, Balasubramanian:2022gmo, Geng:2024jmm}. 

The non-orthogonality of these geometric states is necessary and is universal, since the number of geometric states is continuously infinite while the entropy of a black hole is finite, requiring a huge over-completeness in the geometric states. If $|f\rangle$ with $f$ contained in a finite interval $[f_1,~f_2]$ can span the entire Hilbert space, it is impossible to have a Hermitian operator $\hat{F}$ which satisfies $\hat{F}|f\rangle\approx f|f\rangle$ for all values of $f$. This is because $|f\rangle$ with $f\notin [f_1,~f_2]$ cannot satisfy this approximate equality. Thus, in this case, we can at most impose $\hat{F}|f\rangle\approx f|f\rangle$ for $f\in [f_1,~f_2]$ or for even smaller range of $f$.

In this paper, we construct such Hermitian operators for the length of geodesic timeslice and the timeshift in a two-sided black hole in two-dimensional Jackiew-Teitelboim gravity \cite{Jackiw:1984je, Teitelboim:1983ux, Louis-Martinez:1993bge, Grumiller:2002nm, Almheiri:2014cka, Maldacena:2016hyu, Maldacena:2016upp, Engelsoy:2016xyb, Kitaev:2017awl, Yang:2018gdb, Harlow:2018tqv, Saad:2019lba, Johnson:2019eik}, and show that the spectrum of the both geometric quantities are \emph{discrete}.

\begin{figure}[t]
	\begin{center}
		\includegraphics[width=12cm,clip]{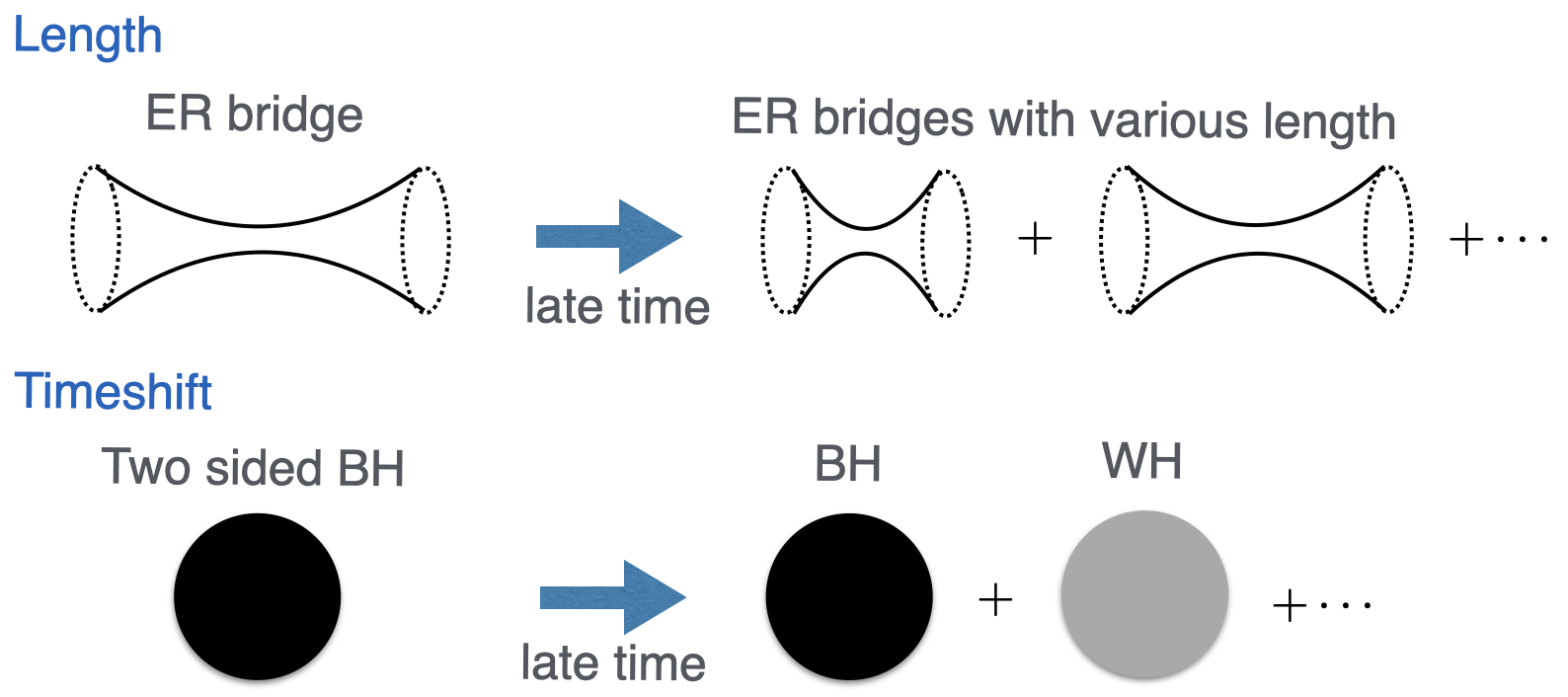}
	\end{center}
	\caption{The main results of this paper. After the Heisenberg time, exponentially long ER bridge becomes an uniform superposition of various length states. Furthermore, a black hole becomes an uniform superposition of BH and WH states.}
	\label{fig:Result}
\end{figure}

The length of the geodesic time slice is identical to the volume of the Einstein-Rosen (ER) bridge, which has been of great interest in terms of the exploration of the black hole interior \cite{Iliesiu:2021ari, Stanford:2022fdt, Iliesiu:2024cnh, Blommaert:2024ftn} as well as relation to the gravity dual of the quantum state complexity \cite{Stanford:2014jda, Brown:2015bva, Brown:2015lvg, Caputa:2017urj, Caputa:2017yrh} and the Fisher information metric \cite{Miyaji:2015woj, Miyaji:2016fse, Belin:2018bpg}. The volume of the ER bridge in a two-sided black hole has been expected to grow linearly in time until it reaches the maximal value of order $e^{S_0}$ from the quantum complexity perspective, which has been confirmed in \cite{Iliesiu:2021ari} for JT gravity. In this paper, we will construct the length operator explicitly and compute the probability distribution of the length \emph{which is completely free from divergence} that was present in previous studies. We find that for an early time, the length probability distribution is sharply peaked at the semiclassical value, and thus the semiclassical approximation is valid. While \emph{at late time, the probability distribution is constant over all possible values of length, implying maximal uncertainty in the length}, see Fig.\ref{fig:Result}. 

We also considered the timeshift at the black hole horizon. This is the canonical conjugate of the two-sided Hamiltonian \cite{Harlow:2018tqv}, which has played an important role in the recent studies on the algebraic structure of the black hole horizon \cite{Leutheusser:2021qhd, Witten:2021unn, Chandrasekaran:2022eqq, Penington:2023dql, Witten:2023xze, Kudler-Flam:2023qfl, Kudler-Flam:2024psh}. As was proposed in \cite{Blommaert:2024ftn}, the timeshift can be used to measure the "age" of the two-sided black hole, capturing the tunneling from the black hole into the white hole via emission of baby universes \cite{Stanford:2022fdt}. In this paper, we construct a non-perturbative timeshift operator. Similar to the length operator, the probability distribution peaked at the semiclassical value at the early time while becoming uniform at the late time. In particular, the probability of finding the white hole and black hole at the late time are equal, showing that the old two-sided black hole is a superposition of black holes and white holes, confirming the idea of "grey hole" proposed in \cite{Susskind:2015toa}.

The organization of this paper is as follows. In section 2, we consider length and timeshift states ignoring the non-orthogonality. In section 3, we explain how discrete spectrum appears and how to construct the non-perturbative length/timeshift operator. In section 4, we apply these non-perturbative operators to understand the late-time interior quantum geometry of the two-sided black hole. We will see that the probability distribution for length becomes flat. We also see that the black and white hole probabilities are equal.

\subsection*{Note Added}
After this paper was finalized, we noticed interesting upcoming work \cite{Akers}, which is based on a distinct construction but contains similar results. It would be very interesting to understand how their construction is related to ours.

%%%%%%%%%%%%%%%%%%%%%%%%%%%%%%%%%%%%%%%%%%%%%%%%%%%%
%%%%%%%%%%%%%%%%%%%%%%%%%%%%%%%%%%%%%%%%%%%%%%%%%%%%
%%%%%%%%%%%%%%%%%%%%%%%%%%%%%%%%%%%%%%%%%%%%%%%%%%%%
%%%%%%%%%%%%%%%%%%%%%%%%%%%%%%%%%%%%%%%%%%%%%%%%%%%%
%%%%%%%%%%%%%%%%%%%%%%%%%%%%%%%%%%%%%%%%%%%%%%%%%%%%

\section{Non-orthogonal Geometric States}

In this section, we will construct non-orthogonal geometric states based on the extended Hartle-Hawking prescription. It turns out the geodesic length state and the timeshift state are special and can be written without referring to the microscopic Hamiltonian.

The bulk single-sided Hamiltonian of JT gravity in the disk topology is
\begin{equation}
    H=\frac{P^2}{2}+2e^{-L},
\end{equation}
where $L$ is the regularized length of the geodesic timeslice, and $P$ is the conjugate momentum of $L$ \cite{Louis-Martinez:1993bge, Harlow:2018tqv}. Since we have $P=i[H,~L]=\dot{L}$, the physical meaning of $P$ is the expansion speed of the length of the geodesic timeslice. Here we set the boundary value of the dilaton as $\phi_b=1$ to simplify notation. Note that the pure JT gravity the energy of the left and right boundaries are equal and Hilbert spaces do not factorize \cite{Harlow:2018tqv}. Thus whenever we denote the energy $E$ in this paper, we will mean by the two-sided energy which is the double of the single-sided one. The density of states at the disk level is
\begin{equation}
\begin{split}
    \rho_{\text{Disk}}(E)=e^{S_0}D_{\text{Disk}}(E)=e^{S_0}\frac{\sinh(2\pi\sqrt{E})}{4\pi^2}.
\end{split}
\end{equation}
The wavefuncitons of $L$ eigenstates can be obtained by solving $-\partial_l^2\psi+4e^{-l}\psi=E\psi$, and one finds
\begin{equation}
\begin{split}
    \psi_{\text{Disk},L}(E,l)&=e^{-S_0/2}2^{3/2}K_{i2\sqrt{E}}(2e^{-l/2}).
\end{split}
\end{equation}
Similarly, the wavefunctions of $P=\dot{L}$ eigenstates are
\begin{equation}
\begin{split}
    \psi_{\text{Disk},P}(E,p)&=e^{-S_0/2}2\Gamma[i(\sqrt{E}-p)]\Gamma[-i(\sqrt{E}+p)].
\end{split}
\end{equation}
In this paper, we will consider the microcanonical ensemble, with a sharp energy window
\begin{equation}
    E\in\Big[E_0-\frac{\Delta E'}{2},~E_0+\frac{\Delta E'}{2}\Big].
\end{equation}
Assuming $\Delta E'$ is small, the total Hilbert space dimension $N'$ is finite and can be approximated as
\begin{equation}
    N'\approx e^{S_0}D_{\text{Disk}}(E_0)\Delta E'.
\end{equation}

%%%%%%%%%%%%%%%%%%%%%%%%%%%%%%%%%%%%%%%%%%%%%%%%%%%%
%%%%%%%%%%%%%%%%%%%%%%%%%%%%%%%%%%%%%%%%%%%%%%%%%%%%
%%%%%%%%%%%%%%%%%%%%%%%%%%%%%%%%%%%%%%%%%%%%%%%%%%%%

\subsection{Geodesic Length States}

Let us denote the microscopic spectrum of the two-sided Hamiltonian as $E_i$. We consider the overlap between the unnormalized thermofield double state $|\text{TFD}(\beta)\rangle=\sum_ie^{-\beta E_i/4}|E_i\rangle$ and the geometric state $|f\rangle$. For some special family of timeslices and associated geometric quantities, the gravitational path-integral for the product of $\langle f|\text{TFD}(\beta)\rangle$s and their conjugates is related to that of the product of $\langle\text{TFD}(\beta')|\text{TFD}(\beta)\rangle=Z((\beta+\beta')/2)$, independently from microscopic Hamiltonians. For example, in the case of the fixed geodesic length timeslice in JT gravity \cite{Penington:2023dql, Iliesiu:2024cnh}, each gravitational configuration with geodesic timeslice can be related to the path-integral with circular AdS boundaries via  
\begin{equation}
    \mathbb{E}\Big[\Big(\int^{\infty}_{-\infty} dl~
    \psi_{\text{Disk},L}(\beta,l)
    \langle l|\text{TFD}(\beta')\rangle\Big)\cdots\Big]=\mathbb{E}\Big[\langle \text{TFD}(\beta)|\text{TFD}(\beta')\rangle\cdots\Big].
\end{equation}
Here $\psi_{\text{Disk},L}(\beta,l)=\int dEe^{S_0}D_{\text{Disk}}(E)e^{-\beta E/4}\psi_{\text{Disk},L}(E,l)$ is the gravitational path-integral in disk topology, with boundary condition composed of fixed geodesic length $l$ and fixed renormalized AdS boundary length $\beta/2$. This constraint is fulfilled when 
\begin{equation}
    |\text{TFD}(\beta)\rangle=\int dl~\psi_{\text{Disk},L}(\beta,l)|l\rangle,
\end{equation}
which can be solved by setting\footnote{It is convenient to use \begin{equation}
	\int_{-\infty}^{\infty}dl~K_{i\sqrt{2E}}(2e^{-l/2})K_{i\sqrt{2E'}}(2e^{-l/2})=\frac{\delta(E-E')}{8D_{\text{Disk}}(E)},
\end{equation}.}
\begin{equation}
    |l\rangle=\sum_{|E_i-E_0|<\Delta E'/2}\psi_{\text{Disk},L}(E_i,l)|E_i\rangle.
\end{equation}
The states for the conjugate momentum $P$ are given by
\begin{equation}
    |p\rangle=\sum_{|E_i-E_0|<\Delta E'/2}\psi_{\text{Disk},P}(E_i,p)|E_i\rangle.
\end{equation}
It is natural to interpret $|l\rangle$ as the eigenvectors of the length operator. However, once we include the effect of baby universe corrections, these states are no longer orthogonal to each other. This implies that the operator $L$ which satisfies $L|l\rangle=l|l\rangle$, cannot be a Hermitian operator and thus is not an observable, i.e. does not give well-defined probability distribution. This issue will be addressed in the next section.

From now on we study properties of the fixed geodesic length wave function $\psi_{\text{Disk},L}(E,l)$ for large $Ee^l$. First, we rewrite the wavefunction as
\begin{equation}
\begin{split}
    \psi_{\text{Disk},L}(E,l)
    &=e^{-S_0/2}2^{3/2}\int_{-\infty}^{\infty}db~\exp\Big(2ib\sqrt{E}-2e^{-l/2}\cosh b\Big).
\end{split}
\end{equation}
The saddle points for this integral are
\begin{equation}
\begin{split}
    b_0&=\pm\arccosh(\sqrt{E}e^{l/2})+\frac{\pi i}{2}(1+2m),~(m\in\mathbb{Z}).
\end{split}
\end{equation}
The integrand around the saddle is, with $b=b_0+\delta b$ and $m=0$,
\begin{equation}
    \exp\Big(-\pi\sqrt{E}\mp 2ie^{-l/2}\sqrt{e^lE-1}\pm 2i\sqrt{E}\arccosh(e^{l/2}\sqrt{E})+ie^{-l/2}\sqrt{e^lE-1}(\delta b)^2\Big).
\end{equation}
\begin{figure}[t]
	\begin{center}
		\includegraphics[width=5cm,clip]{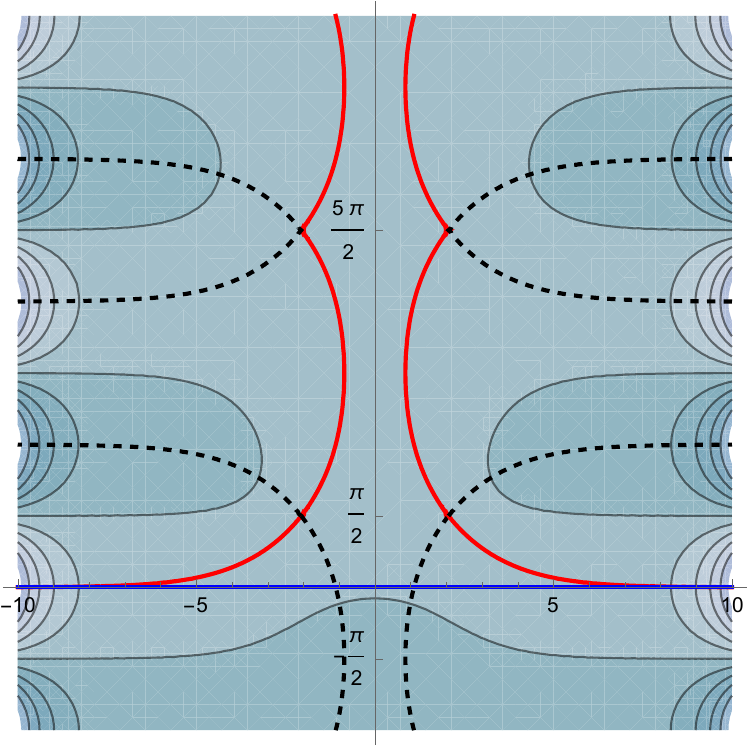}
	\end{center}
	\caption{Change of integration contour in $\psi_{\text{Disk}}(E,l)$ in the complex $b$ plane. Along the dotted black curves and solid red curves, the imaginary part of the integrand is constant. The saddle points for the real part are located at intersections of these lines. We can deform the contour on the real line (solid blue) to the solid red curve. The real part of the integrand (darker blue describes the greater real part) on the red line is maximal at the saddles with $\text{Im}b=\pi/2$, thus they dominate the whole integral at large $\sqrt{E}e^{l/2}$.}
	\label{fig:contour}
\end{figure}
The integration contour can be deformed into a constant phase contour, and the real part of the integrand is maximal at saddle points with $m=0$. Incorporating the Gaussian integral around the saddles, we finally obtain, at large $\sqrt{E}e^{l/2}$,
\begin{equation}
\begin{split}
    \psi_{\text{Disk},L}(E,l)&
    \rightarrow \frac{1}{\sqrt{\pi e^{S_0}D(E_0)}E_0^{1/4}}\cos\left(\sqrt{E}(l+\log 4E - 2)-\frac{\pi}{4}\right).
\end{split}
\end{equation}
Note that the phase comes from the Jacobian $\text{Arg}[\delta b]=\pm\pi/4$. Let us further assume $(t_l-E_0^{-1/2})(\Delta E')^2\ll E_0$, then we can expand inside the exponential and obtain
\begin{equation}
\begin{split}
    \psi_{\text{Disk},L}(E,l)&
    \rightarrow 
    \frac{1}{\sqrt{\pi e^{S_0}D(E_0)}E_0^{1/4}}
    \cos\left(\Big[\sqrt{E_0}(l+\log(4E_0)-2)-2\sqrt{E_0}-\frac{\pi}{4}\Big]+t_l(E-E_0)\right).
\end{split}
\end{equation}
Here we defined 
\begin{equation}
    t_l:=\frac{l+\log (4E_0)}{2\sqrt{E_0}},
\end{equation}
which is the classical relation between the times and the classical length. Later, we will take $t_l=O(e^{S_0}D(E_0))$. Thus the condition is equivalent to
\begin{equation}
    \Delta E'\ll \sqrt{\frac{E_0}{e^{S_0}D(E_0)}},
\end{equation}
therefore the microcanonical window must be sufficiently small. Using the above expression, we obtain the overlap between fixed-length states
\begin{equation}
    \begin{split}
    \langle l|l'\rangle
    &=\sum_{|E_i-E_0|<\Delta E'/2}\psi_{\text{Disk},L}(E_i,l)\psi_{\text{Disk},L}(E_i,l')
    \\&
    \approx
    \int_{E_0-\Delta E'/2}^{E_0+\Delta E'/2}dE~e^{S_0}D(E_1)
    \psi_{\text{Disk},L}(E_i,l)\psi_{\text{Disk},L}(E_i,l')
    \\
    &
    \approx
    \frac{1}{\pi E_0^{1/2}}\Big(\frac{\cos\left(\sqrt{E_0}(l-l')\right)\sin\Big((t_l-t_{l'})\Delta E'/2\Big)}{t_l-t_{l'}}
    \\&
    +
    \frac{\sin\left(\sqrt{E_0}(l+l'+2\log(4E_0)-4)-\frac{\pi}{2}\right)\sin\Big((t_l+t_{l'})\Delta E'/2\Big)}{t_l+t_{l'}}
    \Big).
    \end{split}
\end{equation}
Note that we approximated the first sum over energy eigenstates by the continuous density of states in the second line. The non-orthogonality appearing here is due to the microcanonical ensemble we take. In the canonical ensemble, where we take $\Delta E'=\infty$, we can obtain $\langle l|l'\rangle\underset{\Delta E'\gg 1}{\rightarrow}\frac{\delta(t_l-t_l')}{E_0^{1/2}}$ from the final line, thus the length states are orthogonal at the disk level. Note that such orthogonality at canonical ensemble will break down once we include baby universe corrections, or equivalently microscopic energy statistics. In particular, the normalization of these states is
\begin{equation}
    \begin{split}
    \langle l|l\rangle
    \underset{t_l\gg 1}\rightarrow
    \frac{\Delta E'}{2\pi\sqrt{E_0}},
    \end{split}
\end{equation}
which is independent of $l$.

In the following, we will consider the microcanonical TFD state
\begin{equation}
    |\text{TFD}(t)\rangle=N^{-1/2}\sum_{|E_i-E_0|<\Delta E/2}e^{-iE_it}|E_i\rangle,
\end{equation}
where
\begin{equation}
    N\approx e^{S_0}D_{\text{Disk}}(E_0)\Delta E\approx e^{S_0}\frac{e^{2\pi\sqrt{E}}}{8\pi^2}\Delta E.
\end{equation}
Here we used another microcanonical window which is a subset of the original one, so $\Delta E<\Delta E'$. This state is identical to the fixed $\delta$ state except for the distinct width for the energy window. We quote the result for the overlap of the fixed length states at the disk level with the TFD state. It is given by \cite{Miyaji:2024:}
\begin{equation}\label{eq:TFDlength}
    \begin{split}
    \frac{\langle \text{TFD}(t)| l\rangle\langle  l|\text{TFD}(t)\rangle}{\langle  l| l\rangle}
    &
    \rightarrow
    \frac{2}{\Delta E\Delta E'}\Big[
    \frac{\sin^2\Big((t_l- t)\frac{\Delta E}{2}\Big)}{\Big(t_l- t\Big)^2}
    +\frac{\sin^2\Big((t_l+ t)\frac{\Delta E}{2}\Big)}{\Big(t_l+ t\Big)^2}
    \Big]
    \\&
    +\frac{\pi}{T_H^2\Delta E'}
     \Big[\text{Min}\Big[T_H,~|t_l- t|\Big]+\text{Min}\Big[T_H,~|t_l+t|\Big]\Big].
    \end{split}
\end{equation}
The Heisenberg time $T_H$ is defined by
\begin{equation}
    T_H=\frac{2\pi}{\Delta E'}N'=2\pi e^{S_0}D_{\text{Disk}}(E_0).
\end{equation}
It is natural to interpret $\langle \text{TFD}(t)| l\rangle\langle  l|\text{TFD}(t)\rangle/\langle  l| l\rangle$ as the probability distribution for the length. However $|l\rangle$ are not orthogonal to each other already at the disk level for small $|t_l-t_{l'}|$, and for any $|t_l-t_{l'}|$ once we include baby universe corrections. Thus the distribution does not define a probability distribution. Indeed, (\ref{eq:TFDlength}) cannot be normalized and is constant for any $|t_l|\gg t$. Our strategy in the next section is to define the new fixed geodesic length states and the length operator by projecting out shorter length states from the given length state at the disk level. 

%%%%%%%%%%%%%%%%%%%%%%%%%%%%%%%%%%%%%%%%%%%%%%%%%%%%
%%%%%%%%%%%%%%%%%%%%%%%%%%%%%%%%%%%%%%%%%%%%%%%%%%%%
%%%%%%%%%%%%%%%%%%%%%%%%%%%%%%%%%%%%%%%%%%%%%%%%%%%%
%%%%%%%%%%%%%%%%%%%%%%%%%%%%%%%%%%%%%%%%%%%%%%%%%%%%
%%%%%%%%%%%%%%%%%%%%%%%%%%%%%%%%%%%%%%%%%%%%%%%%%%%%
%%%%%%%%%%%%%%%%%%%%%%%%%%%%%%%%%%%%%%%%%%%%%%%%%%%%

\subsection{Timeshift States}

The bulk one-sided Hamiltonian $H$ is the conjugate momentum of the timeshift $\delta=\frac{t_R+t_L}{2}$. The wavefunction of $\delta$ eigenstates is
\begin{equation}
    \psi_{\text{Disk},\delta}(E,\delta)=\frac{e^{-i\delta E}}{\sqrt{e^{S_0}D_{\text{Disk}}(E)}},
\end{equation}
and the fixed $\delta$ state in the microcanonical ensemble is
\begin{equation}
    |\delta\rangle=\sum_{|E_i-E_0|<\Delta E'/2}\psi_{\text{Disk},\delta}(E_i,\delta)|E_i\rangle,
\end{equation}
which is nothing but the time-evolved microcanonical TFD state which is determined by the data of the energy eigenstates. Thus the product of overlaps can be related to the multi-boundary gravitational path-integral straightforwardly. The overlaps are
\begin{equation}
    \begin{split}
    \langle \delta|\delta'\rangle
    &=\sum_{|E_i-E_0|<\Delta E'/2}
    \psi_{\text{Disk},\delta}(E_i,\delta)^*
    \psi_{\text{Disk},\delta}(E_i,\delta')
    \\&
    =\int_{E_0-\Delta E'/2}^{E_0+\Delta E'/2}dE~e^{S_0}D(E_1)
    \psi_{\text{Disk},\delta}(E_1,\delta)^*
    \psi_{\text{Disk},\delta}(E_1,\delta')
    \\
    &
    =
    \int_{E_0-\Delta E'/2}^{E_0+\Delta E'/2}dE_1~
    e^{i(\delta-\delta')E_1}
    =
    2e^{i(\delta-\delta')E_0}
    \frac{\sin\Big((\delta-\delta')\Delta E'/2\Big)}
    {\delta-\delta'}.    
    \end{split}
\end{equation}
In particular, the normalization of this state is
\begin{equation}
    \begin{split}
    \langle \delta|\delta\rangle=
    \Delta E'.
    \end{split}
\end{equation}
The normalization is independent from $\delta$. 

We note that at the limit of $E_0e^l\gg 1$ and $ \Delta E'\ll \sqrt{\frac{E_0}{e^{S_0}D(E_0)}}$, decomposing the fixed length state into positive and negative $P$ state gives fixed $\delta$ states with $\delta=t_l$ and $\delta=-t_l$. Because $\dot{L}=-i\partial_L$ is the canonical conjugate of $L$ in JT gravity, the exact decomposition of the wavefunction into positive and negative $P$ is
\begin{equation}
\begin{split}
    &\psi^{P>0}_{\text{Disk}}(E,l):=\frac{1}{2\pi}
    \int_0^{\infty} ds~e^{-ils}    
    \int_{-\infty}^{\infty} dl'~e^{il's}\psi_{\text{Disk}}(E,l'),
    \\&
    \psi^{P<0}_{\text{Disk}}(E,l):=\frac{1}{2\pi}
    \int_{-\infty}^0 ds~e^{-ils}    
    \int_{-\infty}^{\infty} dl'~e^{il's}\psi_{\text{Disk}}(E,l').
\end{split}
\end{equation}
They are indeed proportional to fixed $\delta$ wavefunctions at large $Ee^l$, with $\delta=t_l$ for $P>0$ and $\delta=-t_l$ for $P<0$
\begin{equation}
\begin{split}
    &\psi^{P>0}_{\text{Disk}}(E,l)\approx
    \frac{1}{\sqrt{4\pi e^{S_0}D(E_0)}E_0^{1/4}}
    e^{-i\Big[\sqrt{E_0}(l+\log(4E_0)-2)-\frac{\pi}{4}\Big]-it_l(E-E_0)}
    \propto
    \psi_{\text{Disk},\delta}(E,t_l),
    \\&
    \psi^{P<0}_{\text{Disk}}(E,l)\approx
    \frac{1}{\sqrt{4\pi e^{S_0}D(E_0)}E_0^{1/4}}
    e^{i\Big[\sqrt{E_0}(l+\log(4E_0)-2)-\frac{\pi}{4}\Big]+it_l(E-E_0)}\propto\psi_{\text{Disk},\delta}(E,-t_l).
    \end{split}
\end{equation}

Finally, let us review the overlap of these states with the time-evolved thermofield double state. It is given by \cite{Miyaji:2024:}
\begin{equation}\label{eq:TFDtimeshift}
    \begin{split}
    &\frac{\langle \text{TFD}(t)| \delta\rangle\langle  \delta|\text{TFD}(t)\rangle}{\langle  \delta| \delta\rangle}
    \rightarrow
    \frac{4}{\Delta E\Delta E'}
    \frac{\sin^2\Big((\delta- t)\frac{\Delta E}{2}\Big)}{\Big(\delta- t\Big)^2}
    +
    \frac{2\pi}{T_H^2\Delta E'}
     \text{Min}\Big[T_H,~|\delta- t|\Big].
    \end{split}
\end{equation}
Again, we cannot interpret this distribution as a probability distribution for $\delta$, since $|\delta\rangle$ are not orthogonal to each other already at the disk level in the microcanonical ensemble for small $|\delta-\delta'|$, and for arbitrary $|\delta-\delta'|$ due to baby universe corrections. Thus we see that the "time operator" $\hat{\delta}$ conjugate to the Hamiltonian cannot be a Hermitian operator in a finite system. This prohibits us from defining the timeshift operator naively using the fixed timeshift states. Our strategy is to define the new timeshift state by projecting out smaller timeshift states from the given fixed timeshift state at the disk level.

%%%%%%%%%%%%%%%%%%%%%%%%%%%%%%%%%%%%%%%%%%%%%%%%%%%%
%%%%%%%%%%%%%%%%%%%%%%%%%%%%%%%%%%%%%%%%%%%%%%%%%%%%
%%%%%%%%%%%%%%%%%%%%%%%%%%%%%%%%%%%%%%%%%%%%%%%%%%%%
\section{Discrete Spectrum of the Length and the Timeshift}

In the previous section, we have constructed fixed length states and fixed timeshift states in the disk topology. These states are not orthogonal to each other due to the finiteness of the energy window and the baby universe corrections. In order to construct the orthogonal length and timeshift basis, we consider the following positive semi-definite operators for $t_s>0$ and $t_+,~t_->0$ \footnote{More generally, we can consider one parameter family $f_{t_s}(l)$ and $f_{t_s}(\delta)$ of non-negative real functions satisfying $f_{t_s}(l)\leq f_{t_s'}(l)$ and $f_{t_s}(\delta)\leq f_{t_s'}(\delta)$ to consider
\begin{equation}
\begin{split}
    S^L_f[t_0,t_s]&:=\Big(\frac{\Delta E'}{2\pi}\Big)\int_{-\infty}^{\infty} dt_l~f_{t_s}(l)|l\rangle\langle l|,
    \\
    S^\delta_f[-t_-,t_+]&
    =\Big(\frac{\Delta E'}{2\pi}\Big)    
    \int_{-\infty}^{\infty}d\delta~f_{t_s}(\delta)|\delta\rangle\langle \delta|.\nonumber
    \end{split}
\end{equation}
However, the subspaces $H_f^L[t_0,t_s]$ and $H_f^\delta[-t_s,t_s]$ are spanned by $|l\rangle$ with $f_{t_s}(l)\neq 0$ and $|\delta\rangle$ with $f_{t_s}(\delta)\neq0$ respectively. Thus, the precise form of $f_{t_s}$ should not matter in the rest of the paper.
}
\begin{equation}
\begin{split}
    S^L[t_0,t_s]&:=\Big(\frac{\Delta E'}{2\pi}\Big)\int_{t_0}^{t_s} dt_l~|l\rangle\langle l|,
    \\
    S^\delta[-t_-,t_+]&
    :=\Big(\frac{\Delta E'}{2\pi}\Big)    
    \int_{-t_-}^{t_+}d\delta~|\delta\rangle\langle \delta|.
    \end{split}
\end{equation}
We consider the subspace $H^{L}[t_0,t_s]$ and $H^\delta[-t_-,t_+]$ spanned by eigenvectors of $S^L[t_0,t_s]$ and $S^\delta[-t_-,t_+]$ with nonzero eigenvalues of order $O((e^{S_0})^0)$. The projection operators $P^L[t_0,t_s]$ and $P^\delta[-t_-,t_+]$ onto these subspaces can be obtained via the replica trick
\begin{equation}
    \begin{split}
    P^L[t_0,t_s]&=\underset{n\rightarrow +0}{\text{lim}}S^L[t_0,t_s]^{n},\\
    P^\delta[-t_-,t_+]
    &=\underset{n\rightarrow +0}{\text{lim}}S^\delta[-t_-,t_+]^{n},
    \end{split}
\end{equation}
where the replica number $n$ in the right-hand side is analytically continued to positive real number, and then the limit $n\rightarrow +0$ will be taken. Note that we take large $e^{S_0}$ \emph{before} the limit $n\rightarrow 0$, which essentially throws away order $e^{-S_0}$ non-zero eigenvalues in $S^L[t_0,t_s]$ and $S^\delta[-t_-,t_+]$. The Löwner–Heinz inequality \footnote{The Löwner–Heinz inequality tells that for a sequence of operators $A\geq B\geq 0$, we have $A^n\geq B^n\geq 0$ for any $1\geq n\geq 0$. We apply this inequality for positive operators $S^L[t_0,t_s]$ and $S^\delta[-t_-,t_+]$ at $n=0$, which immediately leads to (\ref{eq:inequality}).} tells us that for any $t_s\leq t_s'$ and $t_-\leq t_-',~t_+\leq t_+'$, we have
\begin{equation}\label{eq:inequality}
    \begin{split}
        H^L[t_0,t_s]\subset H^L[t_0,t_s']~~&\text{for}~~t_s\leq t_s'\\
        H^\delta[-t_-,t_+]\subset H^\delta[-t_-',t_+']~~&\text{for}~~t_-\leq t_-',~t_+\leq t_+'.
    \end{split}
\end{equation}
In other words, the supports of $P^L[t_0,t_s]$ and $P^\delta[-t_-,t_+]$ expand monotonically as we increase $t_s$, $t_-$ and $t_+$. 

\begin{figure}[t]
	\begin{center}
		\includegraphics[width=5cm,clip]{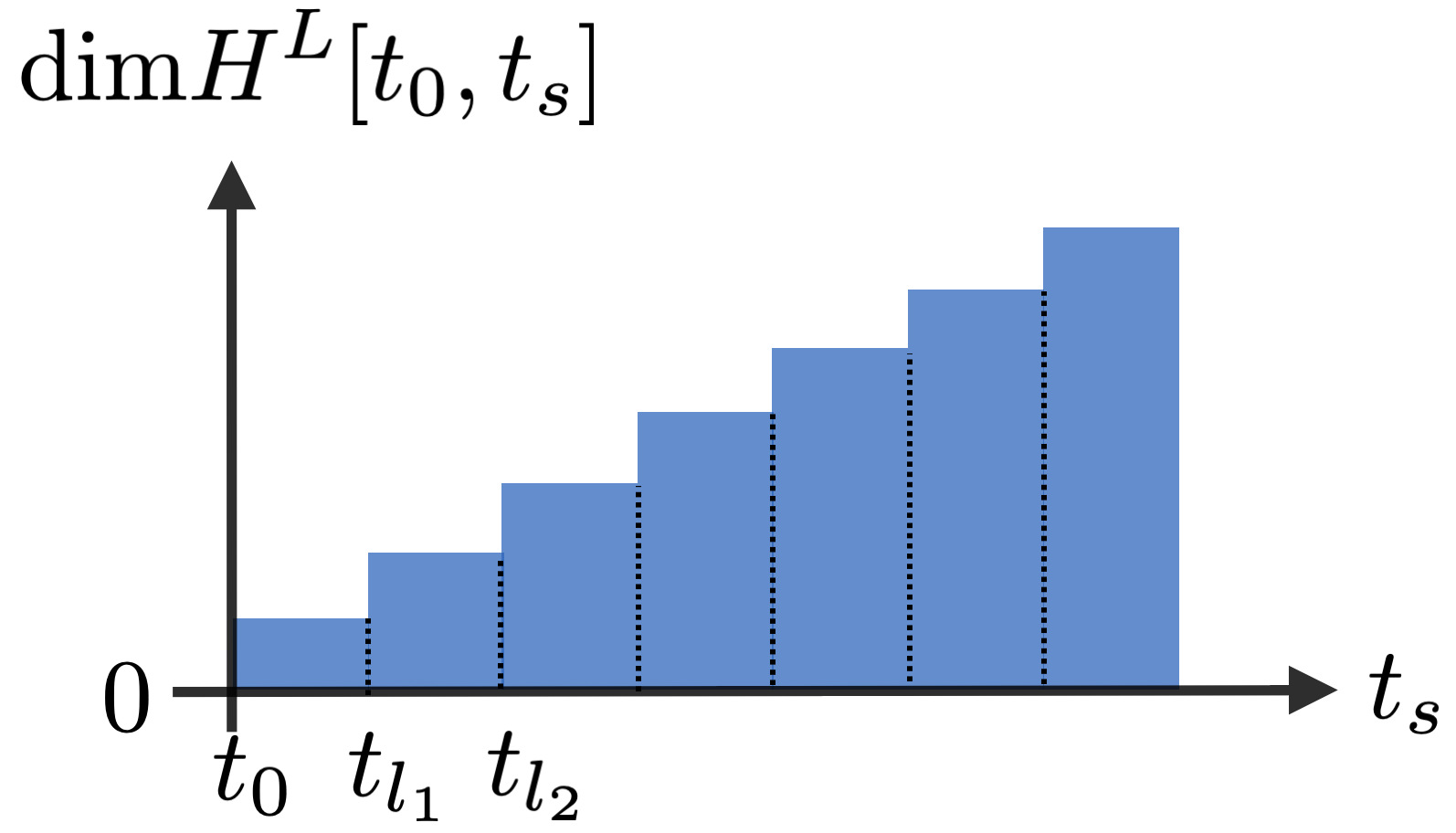}
	\end{center}
	\caption{dim$H^L[t_0,t_s]$ as the function of $t_s$. The discrete jumps determine $t_{l_m}$.}
	\label{fig:dimension}
\end{figure}

Next, let us explain the construction of the orthonormal basis $|\tilde{l}\rangle,~(l_{t_0}\leq l\leq l_{t_s})$ of $H^L[t_0,t_s]$ and $|\tilde{\delta}\rangle,~(-t_-\leq \delta\leq t_+)$ of $H^\delta[-t_-,t_+]$. We focus on the dimension of $H^L[t_0,t_s]$ and $H^\delta[-t_-,t_+]$, which are monotonically increasing discretely as we increase $t_s$, $t_-$ and $t_+$. We denote the values when these discrete jumps happen as $t_{l_1}<t_{l_2}<t_{l_3}<\cdots$ for $t_s$, see Fig. \ref{fig:dimension}. In other words, these values are determined via 
\begin{equation}
    \frac{d}{dt_s}\text{dim} H^L[t_0,t_s]=\sum_{i}\delta(t_s-t_{l_i}),~(t_0<t_{l_1}<t_{l_2}<\cdots).
\end{equation}
At each $t_s=t_{l_i}$, $H^L[t_0,t_s]$ obtains a new normalized state $|l_i\rangle^{NP}$ which is orthogonal to $H^L[t_0,t_s']$ with $t_s'<t_{l_i}$. We call such state as \emph{baby universe corrected length state}.
Similarly, we fix the values of $\delta_{-,i}$ and $\delta_{+,j}$ $(\delta_{+,1}<\delta_{+,2}<\cdots,~\delta_{-,1}<\delta_{-,2}<\cdots)$ via
\begin{equation}
\begin{split}
    &
    \frac{d}{d\delta_+}\text{dim} H^\delta[-\delta_+,\delta_-]\Big{|}_{\delta=\delta_+=\delta_-}=\sum_{i}\delta(\delta-\delta_{+:i}),
    \\&
    \frac{d}{d\delta_-}\text{dim} H^\delta[-\delta_+,\delta_-]\Big{|}_{\delta=\delta_+=\delta_-}=\sum_{j}\delta(\delta+\delta_{-:j}).    
    \end{split}
\end{equation}
Again, for each $\delta=\delta_{+:i}$, $H^\delta[-\delta,\delta]$ obtains a new normalized state $|\delta_{+:i}\rangle^{NP}$ which is orthogonal to $H^\delta[-\delta',\delta']$ with $\delta'<\delta_{+:i}$. The same is true for $\delta_{-:j}$ and the state $|\delta_{-:i}\rangle^{NP}$. We call such states as \emph{baby universe corrected timeshift states}. Thus $H^L[t_0,t_s]$ is spanned by $|l_i\rangle^{NP}$ with $t_0<t_{l_i}<t_s$, and $H^\delta[-\delta,\delta]$ is spanned by
$|\delta_{i}\rangle^{NP}$ with $|\delta_{i}|<\delta$, where we denote both $\delta_{+:i}$ and $\delta_{-:j}$ as $\delta_i$. In other words, we have
\begin{equation}
    \begin{split}
        P^L[t_0,t_s]
        &=\sum_{i:~t_0<t_{l_i}<t_s}|l_i\rangle\langle l_i|^{NP},\\
        P^\delta[-t_s,t_s]
        &
        =\sum_{i:~-t_s<\delta_i<t_s}|\delta_i\rangle\langle \delta_i|^{NP}.    \end{split}
\end{equation}
The new spectrum $l_i$ and $\delta_i$ are interpreted as the baby universe corrected spectrum of the length and the timeshift. Importantly, these spectrums are \emph{discrete}. States $|l_i\rangle$ and $|\delta_i\rangle$ are distinct from the length and timeshift states at the disk level, and form \emph{orthonormal basis} of the Hilbert space \footnote{We note that in general $P^\delta[-t_-,t_+]
        \neq\sum_{i:~-t_-<\delta_i<t_+}|\delta_i\rangle\langle \delta_i|^{NP}$.}. 

We can think that $|l_i\rangle^{NP}$ and $|\delta_i\rangle^{NP}$ are obtained from the Gram-Schmidt orthogonalization procedure of $|l_0\rangle^{\epsilon},~|l_1\rangle^{\epsilon},\cdots$ and $|\delta_0\rangle^{\epsilon},~|\delta_1\rangle^{\epsilon},\cdots$, where $|l\rangle^{\epsilon}\approx|l\rangle$ and $|\delta\rangle^{\epsilon}\approx|\delta\rangle$. Let us consider positive operators for $e^{-S_0}\ll\epsilon\ll 1$
\begin{equation}
\begin{split}
    \Big(\frac{\Delta E'}{2\pi}\Big)\int_{t_{l_i}-\epsilon}^{t_{l_i}} dt_l~|l\rangle\langle l|,~
    \Big(\frac{\Delta E'}{2\pi}\Big)    
    \int_{\delta_{+:i}-\epsilon}^{\delta_{+:i}}d\delta~|\delta\rangle\langle \delta|,
    \end{split}
\end{equation}
such that the subspaces of order one eigenvalue is one-dimensional. We denote the basis state as $|l_i\rangle^{\epsilon}$ and $|\delta_{+:i}\rangle^\epsilon$. These states are not necessarily orthogonal to $H[t_0,t_{l_i}-\epsilon]$ and $H^\delta[-\delta_{+:i}+\epsilon,\delta_{+:i}-\epsilon]$, but are not contained by them. Indeed the states $|l_i\rangle^{NP}$ and $|\delta_i\rangle^{NP}$ can be obtained by extracting components that are orthogonal to these Hilbert spaces. Thus, schematically, we can obtain the baby universe corected states via the following Gram-Schmidt orthogonalization procedure,
\begin{equation}
    \begin{split}
        (|l_0\rangle^\epsilon,~|l_1\rangle^\epsilon,\cdots)&\rightarrow (|l_0\rangle^{NP}=|l_0\rangle^\epsilon,~|l_1\rangle^{NP},\cdots),~(l_i<l_{i+1}),\\
        (|\delta_0\rangle^\epsilon,~|\delta_1\rangle^\epsilon,\cdots)&\rightarrow (|\delta_0\rangle^{NP}=|\delta_0\rangle^\epsilon,~|\delta_1\rangle^{NP},\cdots),~(|\delta_i|<|\delta_{i+1}|).    
        \end{split}
\end{equation}
Here the Gram-Schmidt orthonormalization is performed starting from the reference state $|l_0\rangle$ with $t_{l_1}=t_0$ to a longer length state. In other words, shorter length states are projected out to obtain $|l_i\rangle^{NP}$. For the fixed timeshift state, the orthonormalization is performed from the reference state at $\delta=0$ to larger $|\delta|$.

%%%%%%%%%%%%%%%%%%%%%%%%%%%%%%%%%%%%%%%%%%%%%%%%%%%%
%%%%%%%%%%%%%%%%%%%%%%%%%%%%%%%%%%%%%%%%%%%%%%%%%%%%
%%%%%%%%%%%%%%%%%%%%%%%%%%%%%%%%%%%%%%%%%%%%%%%%%%%%
\subsection{Rank of the Projectors and Density of States}

We evaluate the Renyi version of the rank of operators $S^L[t_0,t_s]$ and $S^\delta[-t_-,t_+]$ via
\begin{equation}
    \begin{split}
    \text{Tr}\Big[S^L[t_0,t_s]^{n}\Big],~
    \text{Tr}\Big[S^\delta[-t_-,t_+]^{n}\Big].
    \end{split}
\end{equation}
To evaluate them, we consider
\begin{equation}
\begin{split}
    Y^L(E_a,E_b:[t_0,t_s])
    &:=
    \langle E_a|S^L[t_0,t_s]|E_b\rangle
    = 
    \int_{t_0}^{t_s} \Big(\frac{\Delta E'}{2\pi}\Big)dt_l~\langle E_a|l\rangle\langle l|E_b\rangle
    \\&
    =\int_{t_0}^{t_s} \Big(\frac{\Delta E'}{2\pi}\Big)dt_l~
    \frac{2^{2}\pi e^{-2\pi\sqrt{E_0}-S_0}}{E_0^{1/2}}
    \Big(
    \cos\left(t_l(E_a-E_b)\right)
    \\&
    +\cos\left(2\Big[\sqrt{E_0}(l+\log(4E_0)-2)-\frac{\pi}{4}\Big]+2t_l(\frac{E_a+E_b}{2}-E_0)\right)
    \Big)
    \\&
    \approx
    \Big(\frac{\Delta E'}{4\pi^2e^{S_0}D(E_0)E_0^{1/2}}\Big)
    \frac{\sin\Big((E_a-E_b)t_s\Big)- \sin\Big((E_a-E_b)t_0\Big)}{E_a-E_b},
    \end{split}
\end{equation}
here we ignored the second term in the integral, since the first terms in the second line is suppressed by $1/E_0$. We also consider
\begin{equation}
\begin{split}
    Y^{\delta}(E_a,E_b:[-t_-,t_+])
    &:=
    \langle E_a|S^\delta[-t_-,t_+]|E_b\rangle
    = 
    \int_{-t_-}^{t_+} \Big(\frac{\Delta E'}{2\pi}\Big)d\delta~\langle E_a|\delta\rangle\langle \delta|E_b\rangle
    \\&
    =
    \Big(\frac{\Delta E'}{\pi e^{S_0}D(E_0)}\Big)
    \frac{e^{i(E_a-E_b)t_-}-e^{-i(E_a-E_b)t_+}}{2i(E_a-E_b)}.
    \end{split}
\end{equation}
Using these quantities, we can express
\begin{equation}
    \begin{split}
    \text{Tr}\Big[S^L[t_0,t_s]^{n}\Big]
    &=\sum_{a_1,\cdots,a_{n+1}}
    \langle E_{a_1}|S^L(t_s)|E_{a_2}\rangle
    \langle E_{a_2}|S^L(t_s)|E_{a_3}\rangle\cdots
    \langle E_{a_{n}}|S^L(t_s)|E_{a_{1}}\rangle\\
    &
    =
    \int^{E_0+\Delta E'/2}_{E_0-\Delta E'/2}~dE_{1}\cdots dE_{n}~
    Y^L(E_1,E_2)Y^L(E_2,E_3)\cdots Y^L(E_n,E_{1})
    \\&\times
    e^{nS_0}
    \langle D(E_1)\cdots D(E_{n})\rangle,
    \end{split}
\end{equation}
For $S^{\delta}$, replace $Y^L$ by $Y^\delta$. We can evaluate these by approximating the density of states correlation function by using the random matrix universality. Namely, when the energy difference is small, we can approximate the density of state correlation function by the sine kernel
\begin{equation}
    K(E_i,E_j)=\frac{\sin\Big(\pi e^{S_0}D(E_0)(E_i-E_j)\Big)}{\pi(E_i-E_j)},
\end{equation}
via the relation from random matrix theory
\begin{equation}
\begin{split}
    &e^{nS_0}\langle D(E_1)\cdots D(E_{n})\rangle
    \\&
    =\sum_{(1,\cdots,N)=(a^{(1)}_1,a^{(1)}_2\cdots)+\cdots}
    \Big(
    \Pi_{i}\delta(E_{a^{(1)}_1},E_{a^{(1)}_2}\cdots) \Big)
    R_{p}(E_{a^{(1)}_1},E_{a^{(2)}_1}\cdots),    
\end{split}
\end{equation}
here the sum is over all possible partition of $(1,\cdots,N)$, counting equivalent partition only once. We defined
\begin{equation}
    \delta(E_1,\cdots,E_m):=\delta(E_1-E_2)\delta(E_2-E_3)\cdots\delta(E_{m-1}-E_m),
\end{equation}
and
\begin{equation}
    R_k(E_1,\cdots,E_k)=\text{det}_{i,j=1,\cdots,k}K(E_i,E_j).
\end{equation}
We also note that we often approximate an integral over rescaled energy difference $u=(E_1-E_2)e^{S_0}$ by an integral over the whole real line, so that the contour integrals can be done conveniently. We thus implicitly assume that $\Delta E'\gg e^{-S_0}$. See also \cite{Iliesiu:2021ari, Miyaji:2024:}.

%%%%%%%%%%%%%%%%%%%%%%%%%%%%%%%%%%%%%%%%%%%%%%%%%%%%
%%%%%%%%%%%%%%%%%%%%%%%%%%%%%%%%%%%%%%%%%%%%%%%%%%%%
\subsubsection*{Short Interval Limit}
Let us consider the case when $t_0,~t_s,~t_-,~t_+\ll T_H$. Then the dominant contribution comes from the disconnected contribution, namely the approximation $\langle D(E_1)\cdots D(E_{n+1})\rangle\approx\langle D(E_1)\rangle\cdots \langle D(E_{n+1})\rangle$ is valid. This contribution is nothing but the semiclassical one. Then we obtain
\footnote{It is convenient to use, for $A,~|B|\geq 0$ with $x_1\neq x_3$ that
\begin{equation}\label{eq:footnote1}
     \int^\infty_{-\infty}dx_2~\frac{e^{iA(x_1-x_2)}-e^{-iB(x_1-x_2)}}{2i(x_1-x_2)}\frac{e^{iA(x_2-x_3)}-e^{-iB(x_2-x_3)}}{2i(x_2-x_3)}=\pi\frac{e^{iA(x_1-x_3)}-e^{-iB(x_1-x_3)}}{2i(x_1-x_3)},
\end{equation}
and for $A,B>0$
\begin{equation}\label{eq:footnote2}
     \int^\infty_{-\infty}dx_2~\frac{\sin(A(x_1-x_2))}{x_1-x_2}\frac{\sin(B(x_2-x_3))}{x_2-x_3}=\pi\frac{\sin(\text{Min}[A,~B](x_1-x_3))}{x_1-x_3}.
\end{equation}
And when $x_1=x_3$, the first expression becomes $\frac{\pi}{2}(A+B)$, and the second expression becomes $\text{Min}[A,~B]$.
}
\begin{equation}\label{eq:densitydisk}
    \begin{split}
        \text{Tr}\Big[S^L[t_0,t_s]^{+0}\Big]
        &=\frac{t_s-t_0}{\pi}\Delta E',
        \\
        \text{Tr}\Big[S^\delta[-t_-,t_+]^{+0}\Big]
        &=
        \frac{t_-+t_+}{2\pi}\Delta E'.
    \end{split}
\end{equation}
This result is remarkable and implies that \emph{the density of geometric states is finite and constant in the microcanonical ensemble}.
%%%%%%%%%%%%%%%%%%%%%%%%%%%%%%%%%%%%%%%%%%%%%%%%%%%%
%%%%%%%%%%%%%%%%%%%%%%%%%%%%%%%%%%%%%%%%%%%%%%%%%%%%
\subsubsection*{First Correction from Finite Width}
We incorporate large $t_0,~t_s,~t_-,~t_+$ correction perturbatively. The correction is given by
\begin{equation}
    \begin{split}
        \text{Tr}\Big[S^L[t_0,t_s]^{n}\Big]^{(1)}
        &=
        e^{2S_0}
        \Big(\frac{\Delta E'}{4 E_0^{1/2}}        
        \Big)^{n-2}
        \frac{n(n-1)}{2}\\&
        \times
        \int^{E_0+\Delta E'/2}_{E_0-\Delta E'/2}~dE_{1}dE_{2}~
        Y^L(E_1,E_2)Y^L(E_2,E_1)
        \langle D(E_1)D(E_2)\rangle_{\text{c}}.
    \end{split}
\end{equation}
For $O^\delta$, replace $Y^L$ by $Y^\delta$. Here we defined
\begin{equation}
    \begin{split}
    \langle D(E_1)D(E_2)\rangle_{\text{c}}
    &:=
    \langle D(E_1)D(E_2)\rangle-\langle D(E_1)\rangle\langle D(E_2)\rangle
    \\&
    \approx e^{-S_0}\delta(E_2-E_3)D_{\text{Disk}}(E_1)-e^{-2S_0}\frac{\sin^2(\pi e^{S_0}D_{\text{Disk}}(E_0)(E_2-E_3))}{\pi^2(E_2-E_3)^2}.
    \end{split}
\end{equation}
We conclude that the first correction vanishes
\begin{equation}
    \begin{split}
        \text{Tr}\Big[S^L[t_0,t_s]^{+0}\Big]^{(1)}
        &=\text{Tr}\Big[S^\delta[-t_-,t_+]^{+0}\Big]^{(1)}=0.
    \end{split}
\end{equation}

%%%%%%%%%%%%%%%%%%%%%%%%%%%%%%%%%%%%%%%%%%%%%%%%%%%%
%%%%%%%%%%%%%%%%%%%%%%%%%%%%%%%%%%%%%%%%%%%%%%%%%%%%
\subsubsection*{Long Interval Limit}

Let us next consider the case when $t_s-t_0,~t_-,~t_+\gg T_H$. Then the dominant contribution comes from the fully connected delta function contribution, namely the approximation $\langle D(E_1)\cdots D(E_{n+1})\rangle\approx e^{-nS_0}D(E_1)\delta(E_1-E_2)\delta(E_2-E_3)\cdots\delta(E_{n}-E_{n+1})$ is valid. Then we can conclude that the rank is maximal
\begin{equation}
    \begin{split}
        \text{Tr}\Big[S^L[t_0,t_s]^{+0}\Big]
        =
        \text{Tr}\Big[S^\delta[-t_-,t_+]^{+0}\Big]
        &=
        e^{S_0}D(E_0)\Delta E'=N'.
    \end{split}
\end{equation}

%%%%%%%%%%%%%%%%%%%%%%%%%%%%%%%%%%%%%%%%%%%%%%%%%%%%
%%%%%%%%%%%%%%%%%%%%%%%%%%%%%%%%%%%%%%%%%%%%%%%%%%%%
\subsubsection*{Density of States}

\begin{figure}[t]
	\begin{center}
		\includegraphics[width=10cm,clip]{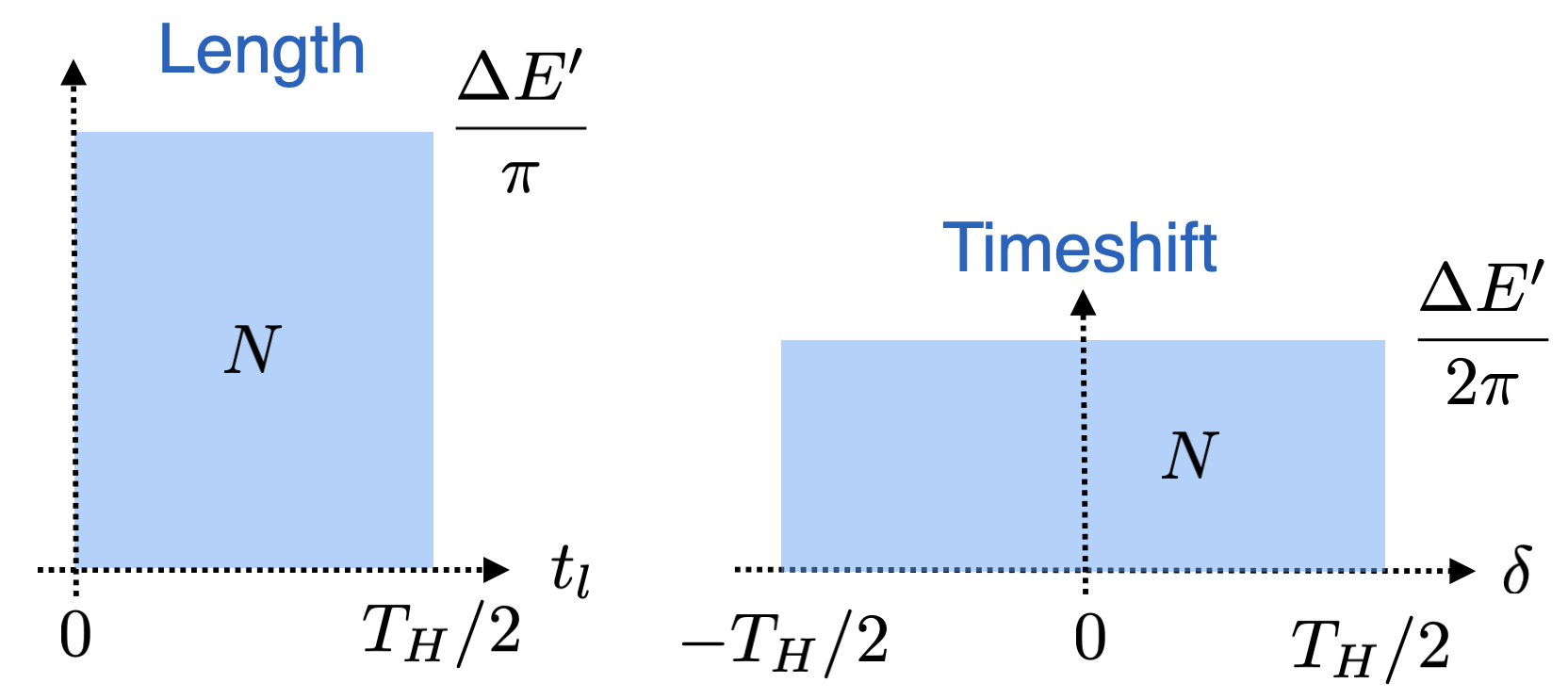}
	\end{center}
	\caption{The density of states of the length $l$ and the timeshift $\delta$ from the estimate \ref{eq:density}. The significant features are that the density of states terminates around $t_l\approx T_H/2$ and $|\delta|\approx T_H/2$, so that the total number of state is given by $N$.}
	\label{fig:densitystate}
\end{figure}

In this paper, we have only computed dominant contributions to the density of geometric states in extreme cases. The absence of the first correction from the finite width effect, suggests that the result at the short interval limit can be interpolated for a finite interval until the total number of states saturates the long interval limit. Assuming this behavior, we estimate that the full density of state, show in Fig. \ref{fig:densitystate},
\begin{equation}\label{eq:density}
    \begin{split}
    \rho(t_s:t_0)&\approx\frac{\Delta E'}{\pi}\theta(\frac{T_H}{2}-t_s),~(t_s>t_0),\\
    \rho([-t_-,t_+])&\approx\frac{\Delta E'}{2\pi}\theta(\frac{T_H}{2}-t_+)\theta(\frac{T_H}{2}-t_-),~(t_-,t_+>0).
    \end{split}
\end{equation}
In other words, \emph{the density of geometric states is finite and constant in the microcanonical ensemble until the total number of states equals the Hilbert space dimension}. The Heaviside theta function in the expression is included in order to be consistent with the long interval limit and to match the Hilbert space dimension. Note that the density of states above depends on what we choose the reference state, such as $t_0$ in $\rho(t_s:t_0)$ and $\delta=0$ in $\rho([-t_-,t_+])$. 

%%%%%%%%%%%%%%%%%%%%%%%%%%%%%%%%%%%%%%%%%%%%%%%%%%%%
%%%%%%%%%%%%%%%%%%%%%%%%%%%%%%%%%%%%%%%%%%%%%%%%%%%%
%%%%%%%%%%%%%%%%%%%%%%%%%%%%%%%%%%%%%%%%%%%%%%%%%%%%

\subsection{Non-perturbative Length and Timeshift Operators}

Using the spectrum and the baby universe corrected geometric states, we can construct non-perturbative geometric operators. The non-perturbative geodesic length operator and timeshift operator can be constructed via
\begin{equation}
    \hat{L}^{\text{NP}}:=\sum_i~l_i|\tilde{l}_i\rangle\langle \tilde{l}_i|,~\hat{\delta}^{\text{NP}}:=\sum_i~\delta_i|\tilde{\delta}_i\rangle\langle \tilde{\delta}_i|.
\end{equation}
These operators indeed approximate the disk level length and timeshift operators for $|l\rangle$ with $t_l\approx t_0$ and $|\delta\rangle$ with small $\delta$, since in that regime we still have $|\tilde{l}\rangle\approx|l\rangle$ and $|\tilde{\delta}\rangle\approx|\delta\rangle$. While for large $l'$ or $|\delta|$, the deviation from the disk level geometric state can be large and such approximation is no longer valid.

%%%%%%%%%%%%%%%%%%%%%%%%%%%%%%%%%%%%%%%%%%%%%%%%%%%%
%%%%%%%%%%%%%%%%%%%%%%%%%%%%%%%%%%%%%%%%%%%%%%%%%%%%
%%%%%%%%%%%%%%%%%%%%%%%%%%%%%%%%%%%%%%%%%%%%%%%%%%%%
%%%%%%%%%%%%%%%%%%%%%%%%%%%%%%%%%%%%%%%%%%%%%%%%%%%%
%%%%%%%%%%%%%%%%%%%%%%%%%%%%%%%%%%%%%%%%%%%%%%%%%%%%
%%%%%%%%%%%%%%%%%%%%%%%%%%%%%%%%%%%%%%%%%%%%%%%%%%%%
\section{Dynamics of the Einstein-Rosen Bridge}

In the following, we use the above non-perturbative length and timeshift operators to analyze the late time behavior of the thermofield double state. For this purpose, it suffices to evaluate the Renyi version of the projection amplitude  
\begin{equation}
    \begin{split}
    Q^L(n,t_0,t_s)&=
    \langle\text{TFD}(t)|S^L(t_0,t_s)^{n}|\text{TFD}(t)\rangle,\\
    Q^\delta(n,t_-,t_+)&=\langle\text{TFD}(t)|S^\delta(t_-,t_+)^{n}|\text{TFD}(t)\rangle,
    \end{split}
\end{equation}
at $n\rightarrow+0$. This projection amplitude is written as
\begin{equation}
    \begin{split}
    Q^L(n,t_0,t_s)
    &=\sum_{a_1,\cdots,a_{n+1}}
    \langle\text{TFD}(t)|E_{a_1}\rangle
    \langle E_{a_1}|S^L(t_s)|E_{a_2}\rangle\cdots
    \langle E_{a_{n}}|S^L(t_s)|E_{a_{n+1}}\rangle\langle E_{a_{n+1}}|\text{TFD}(t)\rangle\\
    &
    =
    \mathcal{N}^{-1}
    \int^{E_0+\Delta E/2}_{E_0-\Delta E/2} dE_1dE_{n+1}\int^{E_0+\Delta E'/2}_{E_0-\Delta E'/2}~dE_{2}\cdots dE_{n}~
    e^{i(E_1-E_{n+1})t}
    \\&
    \times
    Y^L(E_1,E_2)\cdots Y^L(E_n,E_{n+1})
    e^{(n+1)S_0}
    \langle D(E_1)\cdots D(E_{n+1})\rangle.
    \end{split}
\end{equation}
For $Q^\delta$, replace $Y^L$ by $Y^\delta$. The length probability density can then be obtained by taking $t_s$ derivative
\begin{equation}
    \begin{split}
     P^L(t_l)
     &=\frac{d}{dt_s}\Big{|}_{t_s=t_l}Q^L(+0,t_0,t_s).
    \end{split}
\end{equation}
The timeshift probability density is expressed as
\begin{equation}
    \begin{split}
        P^\delta_+(\delta)&
        =\frac{d}{dt_+}\Big{|}_{t_+=t_-=\delta}Q^\delta(+0,t_-,t_+)~(\delta>0),
        \\
        P^\delta_-(\delta)
        &
        =\frac{d}{dt_-}\Big{|}_{t_+=t_-=-\delta}Q^\delta(+0,t_-,t_+)~(\delta<0).
    \end{split}
\end{equation}
Note $t_-=t_+$ is imposed in this expression. We note that when $t_s:=t_-=t_+$ and $t_0=0$, we have $Y^L\propto Y^\delta$. Thus the equality between the length and $|\delta|$ probability holds
\begin{equation}
    Q^\delta(+0,t_s,t_s)^{(0+1)}=Q^L(+0,0,t_s)^{(0+1)}.
\end{equation}

The probability of finding a black hole and a white hole are,
\begin{equation}
    \begin{split}
    P^{\text{BH}}(t)=\int_0^{T_H/2}d\delta'~P^+(\delta'),
    \\
    P^{\text{WH}}(t)=\int_{-T_H/2}^0d\delta'~P^-(\delta').
    \end{split}
\end{equation}

%%%%%%%%%%%%%%%%%%%%%%%%%%%%%%%%%%%%%%%%%%%%%%%%%%%%
%%%%%%%%%%%%%%%%%%%%%%%%%%%%%%%%%%%%%%%%%%%%%%%%%%%%
\subsubsection*{Large $t_l$ and $|\delta|$}

Let us first consider the case when $t_s-t_0,~t_-,~t_+\gg T_H$. Then the dominant contribution comes from the fully connected delta function contribution, namely the approximation $\langle D(E_1)\cdots D(E_{n+1})\rangle\approx e^{-nS_0}D(E_1)\delta(E_1-E_2)\delta(E_2-E_3)\cdots\delta(E_{n}-E_{n+1})$ is valid for $Y^L$, and $\langle D(E_1)\cdots D(E_{n})\rangle\approx e^{-(n-1)S_0}D(E_1)\delta(E_1-E_2)\delta(E_2-E_3)\cdots\delta(E_{n-1}-E_{n})$ is for $Y^{\delta}$. Then we have
\begin{equation}
    \begin{split}
        Q^L(0,t_0,t_s)
        &
        =
        Q^\delta(0,t_-,t_+)
        =
        1,
    \end{split}
\end{equation}
consistent with the probabilistic reasoning.

%%%%%%%%%%%%%%%%%%%%%%%%%%%%%%%%%%%%%%%%%%%%%%%%%%%%
%%%%%%%%%%%%%%%%%%%%%%%%%%%%%%%%%%%%%%%%%%%%%%%%%%%%
%%%%%%%%%%%%%%%%%%%%%%%%%%%%%%%%%%%%%%%%%%%%%%%%%%%%
\subsection{Length of the Einstein-Rosen Bridge}

%%%%%%%%%%%%%%%%%%%%%%%%%%%%%%%%%%%%%%%%%%%%%%%%%%%%
%%%%%%%%%%%%%%%%%%%%%%%%%%%%%%%%%%%%%%%%%%%%%%%%%%%%
\subsubsection*{Small $t_s$}

Let us consider the case when $t_0,t_s\ll T_H$. Then the dominant contribution comes from the disconnected contribution. This is nothing but the semiclassical contribution. Thus we obtain
\begin{equation}
    \begin{split}
        Q^L(0,t_0,t_s)^{(0)}
        &
        =\theta(t+t_s)-\theta(t+t_0)-\theta(t-t_s)+\theta(t-t_0).
    \end{split}
\end{equation}
We thus have
\begin{equation}
    \begin{split}
        P^L(0,t_0,t_s)^{(0)}
        &
        =\delta(t+t_s)+\delta(t-t_s),
    \end{split}
\end{equation}
which matches with the classical length of the Einstein-Rosen bridge.

%%%%%%%%%%%%%%%%%%%%%%%%%%%%%%%%%%%%%%%%%%%%%%%%%%%%
%%%%%%%%%%%%%%%%%%%%%%%%%%%%%%%%%%%%%%%%%%%%%%%%%%%%
\subsubsection*{First Correction from Finite $(t_s-t_0)/T_H$}

\begin{figure}[t]
	\begin{center}
		\includegraphics[width=15cm,clip]{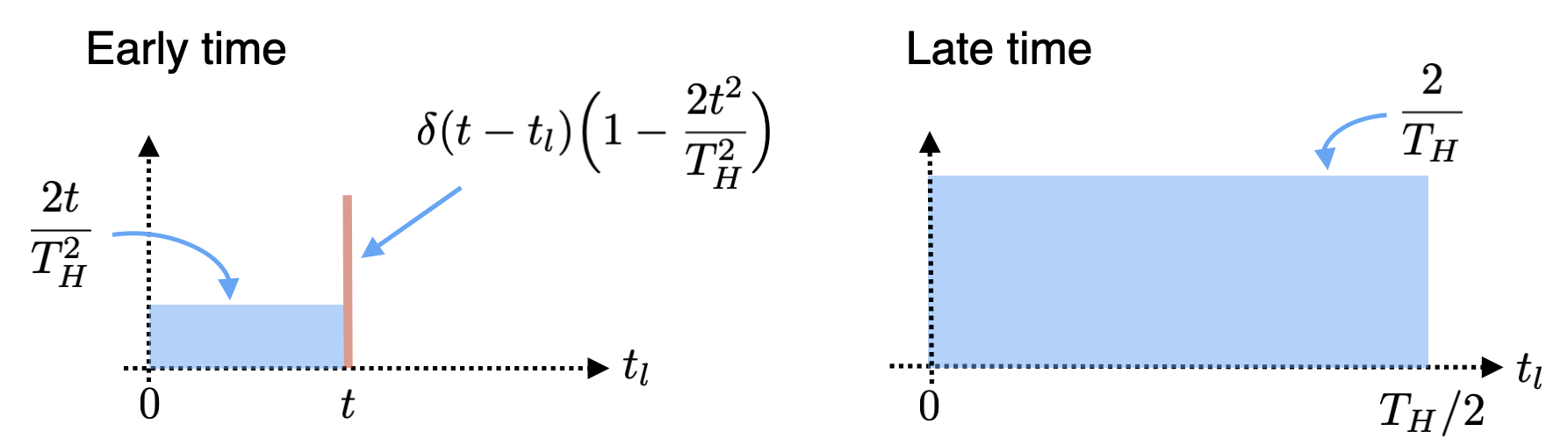}
	\end{center}
	\caption{The probability distribution $P(t_l)$ of the length $l$ of the time-evolved TFD state. The left figure is for early time $t<T_H/2$, and the right figure is for late time $t>3T_H/2$. 
    When $t<T_H/2$, the distribution has a delta function peak at the classical value. From the wormhole effect, the length is distributed over \emph{shorter value uniformly}. When $t>3T_H/2$, the probability is completely uniform.    
    }
	\label{fig:density}
\end{figure}

Next, we study the first correction from $(t_s-t_0)/T_H$ expansion. For this purpose, we include the two-body correlation to the density of state correlation function. We find for $t>0$ and $T_H/2>t_s>t_0>0$ that
\begin{equation}
    \begin{split}
        Q^L(+0,t_0,t_s)^{(1)}
        &
        =
        -
        \Big(\theta(t-t_0)-\theta(t-t_s)\Big)\frac{t^2-2t_0t+t_s^2}{T_H^2}     
        \\&
        +
        \int_{t_0}^{t_s}dt_l~
        \frac{1}{T_H^2}
        \Big(\text{Min}\Big[T_H,~|t_l- t|\Big]+\text{Min}\Big[T_H,~|t_l+t|\Big]\Big)
        \Big].     
    \end{split}
\end{equation}
The detailed derivation is given in appendix \ref{appendix:length}. Let us examine the detailed behavior of $Q^L(+0,t_0,t_s)^{(0)+(1)}$. For simplicity, we approximate as $t_0\approx 0$ in the following. Then we can simplify
\begin{equation}
    \begin{split}
        Q^L(+0,0,t_s)^{(0+1)}
        &=
        1
        +
        \theta(t-t_s)\Big(-1+\frac{2tt_s}{T_H^2}\Big)
        \\&
        +\frac{1}{2T_H^2}
        \Big(
        \theta(t-t_s-T_H)(t-t_s-T_H)^2
        -\theta(t+t_s-T_H)(t+t_s-T_H)^2\Big).
    \end{split}
\end{equation}
The length probability distribution is
\begin{equation}
    \begin{split}
     P(t_l)^{(0+1)}
     &=
     \delta(t-t_l)\Big(1-\frac{2t^2}{T_H^2}\Big)
     +\theta(t-t_l)\frac{2t}{T_H^2}
     \\&
     -\frac{1}{T_H^2}
     \Big(\theta(t-t_l-T_H)(t-t_l-T_H)+\theta(t+t_l-T_H)(t+t_l-T_H)\Big),
    \end{split}
\end{equation}
see Fig. \ref{fig:density}.

%%%%%%%%%%%%%%%%%%%%%%%%%%%%%%%%%%%%%%%%%%%%%%%%%%%%
%%%%%%%%%%%%%%%%%%%%%%%%%%%%%%%%%%%%%%%%%%%%%%%%%%%%
\subsubsection*{Result: Early time $t<T_H/2$:}
The probability distribution is given by
\begin{equation}\label{eq:lengthearly}
    \begin{split}
     P(t_l)^{(0+1)}
     &=
     \delta(t-t_l)\Big(1-\frac{2t^2}{T_H^2}\Big)
     +\theta(t-t_l)\frac{2t}{T_H^2},
    \end{split}
\end{equation}
see the left figure in Fig. \ref{fig:density}.
Thus we have
\begin{equation}
    \begin{split}
        Q^L(+0,0,t_s)^{(0+1)}
        &=
        1-\theta(t-t_s)
        +
        \theta(t-t_s)\frac{2tt_s}{T_H^2}.   
    \end{split}
\end{equation}
Which is $1$ for $t_s>t$, which is consistent with the probabilistic interpretation for $t_s=T_H/2$. Thus in this order, we can see that the baby universe only shortens the classical geodesic time slice, and does not give rise to a longer Einstein-Rosen bridge. The probability of having a shorter ER bridge with $t_l>0$ is uniform. The expectation value of length and its variance are
\begin{equation}
    \mathbb{E}[t_l]^{(0+1)}=t-\frac{t^3}{T_H^2},
    ~\sqrt{\mathbb{E}\Big[(t_l-\mathbb{E}[t_l])^2\Big]^{(0+1)}}=\frac{t^2}{T_H}\sqrt{\frac{2}{3}-\frac{t^2}{T_H^2}}.
\end{equation}
Thus, at the early time, the length expectation value is slightly shorter than the classical value, and the variance is suppressed as $T_H^{-1/2}$. Therefore the usual approximation by classical gravity is still valid.

%%%%%%%%%%%%%%%%%%%%%%%%%%%%%%%%%%%%%%%%%%%%%%%%%%%%
%%%%%%%%%%%%%%%%%%%%%%%%%%%%%%%%%%%%%%%%%%%%%%%%%%%%
\subsubsection*{Result: Late time $T_H/2<t$:}
The probability distribution is given by
\begin{equation}
    \begin{split}
     P(t_l)^{(0+1)}
     &=
     \frac{2t}{T_H^2}
     -\frac{1}{T_H^2}
     \Big(\theta(t-t_l-T_H)(t-t_l-T_H)+\theta(t+t_l-T_H)(t+t_l-T_H)\Big).
    \end{split}
\end{equation}
Let us focus on the case $3T_H/2<t$. Then we have
\begin{equation}
    \begin{split}
        P(t_l)^{(0+1)}=\frac{2}{T_H},~Q^L(+0,0,t_s)^{(0+1)}=\frac{2t_s}{T_H},
    \end{split}
\end{equation}
see the right figure in Fig. \ref{fig:density}.
$Q^L(+0,0,t_s)^{(0+1)}$ becomes $1$ when $t_s=T_H/2$, and the density of length is uniform. The expectation value of length and its variance are
\begin{equation}
    \mathbb{E}[t_l]^{(0+1)}=\frac{T_H}{4},
    ~\sqrt{\mathbb{E}\Big[(t_l-\mathbb{E}[t_l])^2\Big]^{(0+1)}}=\frac{T_H}{4\sqrt{3}}.
\end{equation}
Thus the length expectation value is independent from time $t$ and is significantly shorter than the classical value. The variance is of the same order as the length expectation value. To account for the intermediate time $T_H/2<t<3T_H/2$, the current leading order calculation is not sufficient and we need higher order calculations.

%%%%%%%%%%%%%%%%%%%%%%%%%%%%%%%%%%%%%%%%%%%%%%%%%%%%
%%%%%%%%%%%%%%%%%%%%%%%%%%%%%%%%%%%%%%%%%%%%%%%%%%%%
%%%%%%%%%%%%%%%%%%%%%%%%%%%%%%%%%%%%%%%%%%%%%%%%%%%%
\subsection{Timeshift at the Black Hole Horizon}

%%%%%%%%%%%%%%%%%%%%%%%%%%%%%%%%%%%%%%%%%%%%%%%%%%%%
%%%%%%%%%%%%%%%%%%%%%%%%%%%%%%%%%%%%%%%%%%%%%%%%%%%%
\subsubsection*{Small $t_-,~t_+$}

Let us consider the case when $t_-,~t_+\ll T_H$. Then the dominant contribution comes from the disconnected contribution. Thus
\begin{equation}
    \begin{split}
        Q^\delta(0,t_-,t_+)^{(0)}
        &=
        \theta(t+t_-)-\theta(t-t_+).
    \end{split}
\end{equation}
We thus have
\begin{equation}
    \begin{split}
        P^\delta_+(0,\delta)^{(0)}
        =\delta(t-\delta),~P^\delta_-(0,\delta)^{(0)}=\delta(t+\delta).
    \end{split}
\end{equation}
Again we reproduce the classical expectation.

%%%%%%%%%%%%%%%%%%%%%%%%%%%%%%%%%%%%%%%%%%%%%%%%%%%%
%%%%%%%%%%%%%%%%%%%%%%%%%%%%%%%%%%%%%%%%%%%%%%%%%%%%
\subsubsection*{First Correction from Finite $(t_s-t_0)/T_H$}

The first correction from $(t_s-t_0)/T_H$ expansion for $Q(+0,t,t_-,t+)$ is, for $0<t,t_-,t_+$ and $t_-+t_+<T_H$,
\begin{equation}
    \begin{split}
        Q^\delta(+0,t_-,t_+)^{(1)}
        &
        =
        -\theta(-t+t_+)        
        \Big(t^2+t(t_--t_+)+\frac{t_-^2+t_+^2}{2}
        \Big)\frac{1}{T_H^2}
        \\&
        +
        \int_{-t_-}^{t_+}d\delta~\frac{1}{T_H^2}
        \text{Min}\Big[T_H,~|\delta- t|\Big].     
    \end{split}
\end{equation}
see appendix \ref{appendix:timeshift} for more detailed calculations. Let us examine the detailed behavior of $Q^\delta(+0,t_-,t_+)^{(0)+(1)}$. The probability distribution for $\delta$ is
\begin{equation}
    \begin{split}
        P^+(\delta)^{(0+1)}&
        =\delta(t-\delta)\Big(1-\frac{2t^2}{T_H^2}\Big)
        -\theta(\delta-t)        
        \frac{\delta-t}{T_H^2}
        +
        \frac{1}{T_H^2}
        \text{Min}\Big[T_H,~|\delta- t|\Big],
        \\
        P^-(\delta)^{(0+1)}
        &
        =-\theta(-t-\delta)        
        \frac{t-\delta}{T_H^2}
        +
        \frac{1}{T_H^2}
        \text{Min}\Big[T_H,~|\delta-t|\Big],
    \end{split}
\end{equation}
see Fig. \ref{fig:densitytimeshift}.

%%%%%%%%%%%%%%%%%%%%%%%%%%%%%%%%%%%%%%%%%%%%%%%%%%%%
%%%%%%%%%%%%%%%%%%%%%%%%%%%%%%%%%%%%%%%%%%%%%%%%%%%%
\subsubsection*{Result: Early time $t<T_H/2$:}
The probability distribution is
\begin{equation}\label{eq:timeshiftearly}
    \begin{split}
        P^+(\delta)^{(0+1)}&
        =\delta(t-\delta)\Big(1-\frac{2t^2}{T_H^2}\Big)
        +
        \frac{1}{T_H^2}\theta(t-\delta)(t-\delta),
        \\
        P^-(\delta)^{(0+1)}
        &
        =\theta(t+\delta)        
        \frac{t-\delta}{T_H^2},
    \end{split}
\end{equation}
see the left figure of Fig. \ref{fig:densitytimeshift}.
The integrated probability, which corresponds to the BH and WH respectively, is
\begin{equation}\label{eq:whprobabilityearly}
    \begin{split}
    P^{\text{BH}}(t)
    =1-\frac{3t^2}{2T_H^2},~
    P^{\text{WH}}(t)
    =\frac{3t^2}{2T_H^2}.
    \end{split}
\end{equation}
The expectation value of the timeshift and its variance are
\begin{equation}
    \mathbb{E}[\delta]^{(0+1)}=t\Big(1-\frac{8t^2}{3T_H^2}\Big),
    ~\sqrt{\mathbb{E}\Big[(\delta-\mathbb{E}[\delta])^2\Big]^{(0+1)}}=\frac{2t^2}{T_H}\sqrt{1+\frac{16t^2}{3T_H^2}}.
\end{equation}

%%%%%%%%%%%%%%%%%%%%%%%%%%%%%%%%%%%%%%%%%%%%%%%%%%%%
%%%%%%%%%%%%%%%%%%%%%%%%%%%%%%%%%%%%%%%%%%%%%%%%%%%%
\subsubsection*{Result: Late time $t>T_H/2$:}
\begin{figure}[t]
	\begin{center}
		\includegraphics[width=15cm,clip]{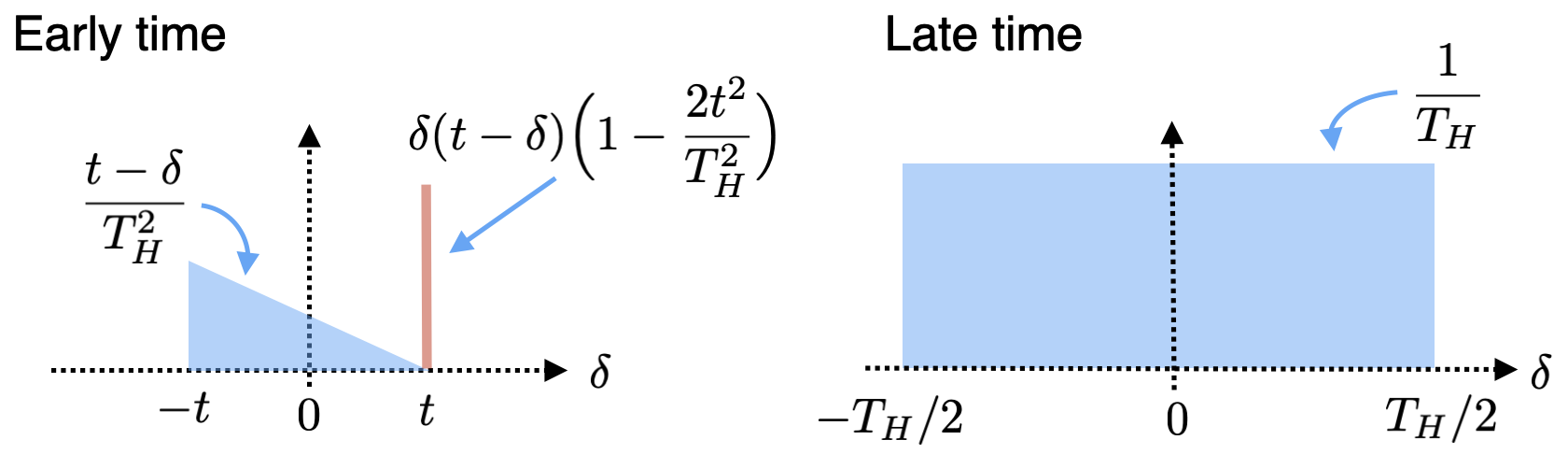}
	\end{center}
	\caption{The probability distribution $P(\delta)$ of the timeshift $\delta$ of the time-evolved TFD state. The left figure is for early time $t<T_H/2$, and the right figure is for late time $t>3T_H/2$. When $t<T_H/2$, the distribution has a delta function peak at the classical value, and the timeshift is linearly distributed for $|\delta|<t$. When $t>3T_H/2$, the probability distribution becomes uniform.}
	\label{fig:densitytimeshift}
\end{figure}
The probability distribution is
\begin{equation}
    \begin{split}
        P^+(\delta)^{(0+1)}&
        =
        \frac{1}{T_H^2}
        \text{Min}\Big[T_H,~|\delta- t|\Big],
        \\
        P^-(\delta)^{(0+1)}
        &
        =
        \frac{1}{T_H^2}
        \text{Min}\Big[T_H,~|\delta-t|\Big].
    \end{split}
\end{equation}
When $t>3T_H/2$, then we have
\begin{equation}
    \begin{split}
        P^+(\delta)^{(0+1)}=
        P^-(\delta)^{(0+1)}=\frac{1}{T_H},
    \end{split}
\end{equation}
see the right figure of Fig. \ref{fig:densitytimeshift}.
Thus we have
\begin{equation}
    \begin{split}
    P^{\text{BH}}(t)
    =
    P^{\text{WH}}(t)
    =\frac{1}{2},
    \end{split}
\end{equation}
indicating equal probability between a black hole and a white hole at the late time. The expectation value of the timeshift and its variance are
\begin{equation}
    \mathbb{E}[\delta]^{(0+1)}=0,
    ~\sqrt{\mathbb{E}\Big[(\delta-\mathbb{E}[\delta])^2\Big]^{(0+1)}}=\frac{T_H}{2\sqrt{3}}.
\end{equation}
Again, to account for the intermediate time $T_H/2<t<3T_H/2$, we need higher order calculations.  

%%%%%%%%%%%%%%%%%%%%%%%%%%%%%%%%%%%%%%%%%%%%%%%%%%%%
%%%%%%%%%%%%%%%%%%%%%%%%%%%%%%%%%%%%%%%%%%%%%%%%%%%%
%%%%%%%%%%%%%%%%%%%%%%%%%%%%%%%%%%%%%%%%%%%%%%%%%%%%
%%%%%%%%%%%%%%%%%%%%%%%%%%%%%%%%%%%%%%%%%%%%%%%%%%%%
\section{Discussions}

In this paper, we constructed non-perturbative length and timeshift operators. It turned out that their spectrum is discrete in accord with the dimensionality of the microcanonical Hilbert space. By considering the timeshift operator, we confirmed that the late-time TFD state is the mixture of the black hole and white hole states with equal probability. The most urgent future direction is to fully incorporate the finite $(t_s-t_0)/T_H$ and $t_s/T_H$ effects into the density of states and the probability distribution, in order to justify our results at late time. 
We expect the construction described in this paper can be applied to general bulk geometric quantities. Furthermore, we can analyze the \emph{discrete} version of the algebraic structure at the horizon as well as the density of geometric states in various backgrounds including de Sitter space. It would also be interesting to understand the relation to approaches to Hartle-Hawking wavefunction from Liouville action \cite{Caputa:2017urj, Caputa:2017yrh, Boruch:2020wax, Boruch:2021hqs} and $T\bar{T}$ deformation \cite{Araujo-Regado:2022gvw}. We also emphasize the close relations to the constructions of dual states in \cite{Miyaji:2015fia, Doi:2024nty}. We conclude with further comments and future directions.

%%%%%%%%%%%%%%%%%%%%%%%%%%%%%%%%%%%%%%%%%%%%%%%%%%%%
%%%%%%%%%%%%%%%%%%%%%%%%%%%%%%%%%%%%%%%%%%%%%%%%%%%%
%%%%%%%%%%%%%%%%%%%%%%%%%%%%%%%%%%%%%%%%%%%%%%%%%%%%
\subsection*{Relation to Other Definitions of Length}

In \cite{Stanford:2022fdt}, the minimal length geodesic timeslice was considered to define the probability distribution as well as transition probability into a white hole. Although the current approach is similar to theirs, the final results are somewhat different; we can compare the results for probability distribution (\ref{eq:lengthearly}), (\ref{eq:timeshiftearly}), and the white hole probability (\ref{eq:whprobabilityearly}). Our results indicate that transition into shorter wormholes has constant probability in $t_l$ as can be seen in (\ref{eq:lengthearly}), (\ref{eq:timeshiftearly}) and in (\ref{eq:whprobabilityearly}), while in \cite{Stanford:2022fdt}, the probability is proportional to $(t-t_l)$. It is very interesting to understand the origin of this difference and give a detailed geometric understanding of our approach. Note that there is no contradiction here, as the two approaches are not equivalent. It is also intriguing to understand the connection to the minimal geodesic cut considered in \cite{Penington:2023dql}. 

The nice feature of the current construction of a non-perturbative geometric operator is that it is free from divergence. The divergence can come from the constant overlap with arbitrary long length and timeshift state as displayed in (\ref{eq:TFDlength}) and (\ref{eq:TFDtimeshift}), which leads to divergence of the total probability and thus expectation value of geometric quantities\footnote{Note that neither the length state nor the timeshift state give the resolution of identity. As we can see by integrating over $l$ or $\delta$ of (\ref{eq:TFDlength}) and (\ref{eq:TFDtimeshift}), we can see
\begin{equation}\label{eq:diagonal}
    \int_{-\infty}^{\infty}dl~|l\rangle\langle l|\neq\mathbb{I},~\int_{-\infty}^{\infty}d\delta~|\delta\rangle\langle \delta|\neq 2\pi\mathbb{I},\nonumber
\end{equation}
because the left-hand sides of the two equations are \emph{divergent} for any time-evolved TFD state. The equalities (\ref{eq:diagonal}) hold at the disk level, but fail badly once we include the baby universe corrections. Note that, unlike many other cases, the disk answer is not even close to the ensemble-averaged answer.}. In our case the spectrum of the length and the timeshift are finite, and the total probability is normalized, therefore such divergence does not arise.

%%%%%%%%%%%%%%%%%%%%%%%%%%%%%%%%%%%%%%%%%%%%%%%%%%%%
%%%%%%%%%%%%%%%%%%%%%%%%%%%%%%%%%%%%%%%%%%%%%%%%%%%%
%%%%%%%%%%%%%%%%%%%%%%%%%%%%%%%%%%%%%%%%%%%%%%%%%%%%
\subsection*{Dependence on the Reference State}

The non-perturbative length operator depends on the reference state in order to construct the basis of geometric states from that reference state. The most remarkable feature is that geometric interpretation can drastically depend on the choice of the basis. If we consider the short ER bridge basis, then at a sufficiently late time, the overlap with the TFD state spread homogeneously over the basis and there is no classical fixed length timeslice at all, as we saw in the main text. On the other hand, if we consider the long ER bridge basis, with length of order the Heisenberg time, then there was no classical fixed length timeslice at early times, yet at some point in the late time, classical geometry suddenly emerges out of nowhere. Of course, such sudden pop-up at fixed time is due to the fine-tuning of the initial state; for generic initial state, we expect it takes Poincare recurrence time to have the consentration of the wavefunction in length basis. For the problem of black hole vs white hole, in the short ER bridge basis, the probability of having a white hole is small at an early time and becomes $1/2$ at a late time. On the other hand, on the long ER bridge basis, the probability of finding a white hole is $1/2$ at an early time and becomes small at some point in the late time. It would be very interesting to understand the bulk physical implications of such reference dependence. 
%%%%%%%%%%%%%%%%%%%%%%%%%%%%%%%%%%%%%%%%%%%%%%%%%%%%
%%%%%%%%%%%%%%%%%%%%%%%%%%%%%%%%%%%%%%%%%%%%%%%%%%%%
%%%%%%%%%%%%%%%%%%%%%%%%%%%%%%%%%%%%%%%%%%%%%%%%%%%%
\subsection*{Probes for Generic Quantum Systems}

The construction of our non-perturbative length and the timeshift operator has a close resemblance to the spread complexity \cite{Balasubramanian:2022tpr, Balasubramanian:2023kwd, Caputa:2024vrn, Nandy:2024htc}. Indeed, for early time in double-scaled SYK model, the length becomes identical to the spread complexity \cite{Lin:2022rbf, Rabinovici:2023yex}. In the spread complexity, one acts Hamiltonian to the initial state multiple times to create a set of states and applies the Gram-Schmidt orthogonalization to these states to construct a basis for the (sub)space. In the current case, we instead fix the reference state and construct projectors onto $O(1)$ eigenvalues of positive semi-definite operators $S^L[t_0,t_s]$ and $S^\delta[-t_-,t_+]$. More generally, we can consider the subspace spanned by $O(1)$ eigenvalue eigenstates of the time-averaged density matrix
\begin{equation}
    \bar{\rho}[0,T]:=\frac{1}{T}\int_{0}^Tdt~\rho(t),
\end{equation}
and construct the orthonormal basis $|T_i\rangle^{NP}$ of the Hilbert space. Using that basis, we can construct the "time" operator
\begin{equation}
    \hat{T}:=\sum_iT_i|T_i\rangle\langle T_i|^{NP}.
\end{equation}
It would be very interesting to explore various relations further and compute these quantities in generic quantum systems.

%%%%%%%%%%%%%%%%%%%%%%%%%%%%%%%%%%%%%%%%%%%%%%%%%%%%
%%%%%%%%%%%%%%%%%%%%%%%%%%%%%%%%%%%%%%%%%%%%%%%%%%%%
%%%%%%%%%%%%%%%%%%%%%%%%%%%%%%%%%%%%%%%%%%%%%%%%%%%%
%%%%%%%%%%%%%%%%%%%%%%%%%%%%%%%%%%%%%%%%%%%%%%%%%%%%
%%%%%%%%%%%%%%%%%%%%%%%%%%%%%%%%%%%%%%%%%%%%%%%%%%%%
%%%%%%%%%%%%%%%%%%%%%%%%%%%%%%%%%%%%%%%%%%%%%%%%%%%%
%%%%%%%%%%%%%%%%%%%%%%%%%%%%%%%%%%%%%%%%%%%%%%%%%%%%
%%%%%%%%%%%%%%%%%%%%%%%%%%%%%%%%%%%%%%%%%%%%%%%%%%%%

\acknowledgments

We are grateful to Chris Akers, Vijay Balasubramanian, Hugo Camargo, Luca Iliesiu, Don Marolf, Kazumi Okuyama, Shanming Ruan, Phil Saad, Tadashi Takayanagi and Zhenbin Yang for the discussions. We thank Soichiro Mori, Shono Shibuya, Kotaro Shinmyo, and Kazuyoshi Yano for various conversations. MM is supported by JSPS KAKENHI Grant-in-Aid for Early-Career Scientists (24K17044). This research was supported in part by the International Centre for Theoretical Sciences (ICTS) for the program - Quantum Information, Quantum Field Theory and Gravity (code: ICTS/qftg2024/08).

%%%%%%%%%%%%%%%%%%%%%%%%%%%%%%%%%%%%%%%%%%%%%%%%%%
%%%%%%%%%%%%%%%%%%%%%%%%%%%%%%%%%%%%%%%%%%%%%%%%%%%%
%%%%%%%%%%%%%%%%%%%%%%%%%%%%%%%%%%%%%%%%%%%%%%%%%%%%
%%%%%%%%%%%%%%%%%%%%%%%%%%%%%%%%%%%%%%%%%%%%%%%%%%%%
%%%%%%%%%%%%%%%%%%%%%%%%%%%%%%%%%%%%%%%%%%%%%%%%%%%%
%%%%%%%%%%%%%%%%%%%%%%%%%%%%%%%%%%%%%%%%%%%%%%%%%%
%%%%%%%%%%%%%%%%%%%%%%%%%%%%%%%%%%%%%%%%%%%%%%%%%%%%
%%%%%%%%%%%%%%%%%%%%%%%%%%%%%%%%%%%%%%%%%%%%%%%%%%%%
%%%%%%%%%%%%%%%%%%%%%%%%%%%%%%%%%%%%%%%%%%%%%%%%%%%%
%%%%%%%%%%%%%%%%%%%%%%%%%%%%%%%%%%%%%%%%%%%%%%%%%%%%

\appendix

%%%%%%%%%%%%%%%%%%%%%%%%%%%%%%%%%%%%%%%%%%%%%%%%%%
%%%%%%%%%%%%%%%%%%%%%%%%%%%%%%%%%%%%%%%%%%%%%%%%%%%%
%%%%%%%%%%%%%%%%%%%%%%%%%%%%%%%%%%%%%%%%%%%%%%%%%%%%
%%%%%%%%%%%%%%%%%%%%%%%%%%%%%%%%%%%%%%%%%%%%%%%%%%%%
%%%%%%%%%%%%%%%%%%%%%%%%%%%%%%%%%%%%%%%%%%%%%%%%%%%%
%%%%%%%%%%%%%%%%%%%%%%%%%%%%%%%%%%%%%%%%%%%%%%%%%%%%
%%%%%%%%%%%%%%%%%%%%%%%%%%%%%%%%%%%%%%%%%%%%%%%%%%%%
%%%%%%%%%%%%%%%%%%%%%%%%%%%%%%%%%%%%%%%%%%%%%%%%%%%%
%%%%%%%%%%%%%%%%%%%%%%%%%%%%%%%%%%%%%%%%%%%%%%%%%%%%
%%%%%%%%%%%%%%%%%%%%%%%%%%%%%%%%%%%%%%%%%%%%%%%%%%%%

\section{Detailed Calculation of the Corrections}
%%%%%%%%%%%%%%%%%%%%%%%%%%%%%%%%%%%%%%%%%%%%%%%%%%%%
%%%%%%%%%%%%%%%%%%%%%%%%%%%%%%%%%%%%%%%%%%%%%%%%%%%%
%%%%%%%%%%%%%%%%%%%%%%%%%%%%%%%%%%%%%%%%%%%%%%%%%%%%

\subsection{Length} \label{appendix:length}

We present detailed calculations of finite $(t_s-t_0)/T_H$ correction to the $Q^L$ and $O^L$. The correction is
\begin{equation}
    \begin{split}
        Q^L(n,t_k)^{(1)}
        &=
        \mathcal{N}^{-1}e^{4S_0}
        \Big(\frac{\Delta E'}{4\pi E_0^{1/2}}        
        \Big)^{n-3}
        \int^{E_0+\Delta E/2}_{E_0-\Delta E/2} dE_1dE_{n+1}\int^{E_0+\Delta E'/2}_{E_0-\Delta E'/2}~dE_{2}dE_{3}~
        e^{i(E_1-E_{n+1})t}
        \\&
        \times
        \frac{(n-1)(n-2)}{2}
        Y(E_1,E_2)Y(E_2,E_3)Y(E_3,E_{n+1})
        D(E_1)\langle D(E_2)D(E_3)\rangle_{\text{c}}D(E_{n+1})    
        \\&
        +
        \mathcal{N}^{-1}
        e^{3S_0}
        \Big(\frac{\Delta E'}{4\pi E_0^{1/2}}        
        \Big)^{n-2}
        \int^{E_0+\Delta E/2}_{E_0-\Delta E/2} dE_1dE_{n+1}\int^{E_0+\Delta E'/2}_{E_0-\Delta E'/2}~dE_{2}~
        e^{i(E_1-E_{n+1})t}
        \\&
        \times
        (n-1)Y(E_1,E_2)Y(E_2,E_{n+1})
        \Big(\langle D(E_1)D(E_2)\rangle_{\text{c}}D(E_{n+1})+D(E_1)\langle D(E_2)D(E_{n+1})\rangle_{\text{c}}
        \Big) 
        \\&
        +
        \mathcal{N}^{-1}
        e^{2S_0}
        \Big(\frac{\Delta E'}{4\pi E_0^{1/2}}        
        \Big)^{n-1}
        \int^{E_0+\Delta E/2}_{E_0-\Delta E/2} dE_1dE_{n+1}~
        e^{i(E_1-E_{n+1})t}
        Y(E_1,E_{n+1})
        \langle D(E_1)D(E_{n+1})\rangle_{\text{c}}.
    \end{split}
\end{equation}
We assume $t>0$. The first term is
\begin{equation}
    \begin{split}
        &\mathcal{N}^{-1}e^{S_0}
        \Big(\frac{\Delta E'}{4\pi E_0^{1/2}}        
        \Big)^{n-1}
        \int^{E_0+\Delta E/2}_{E_0-\Delta E/2} dE_1dE_{n+1}~
        e^{i(E_1-E_{n+1})t}
        \frac{(n-1)(n-2)}{2}
        Y(E_1,E_{n+1})
        D(E_{n+1})\frac{t_s-t_0}{\pi}
        \\&-
        \mathcal{N}^{-1}e^{2S_0}
        \Big(\frac{\Delta E'}{4\pi E_0^{1/2}}        
        \Big)^{n-3}
        \int^{E_0+\Delta E/2}_{E_0-\Delta E/2} dE_1dE_{n+1}\int^{E_0+\Delta E'/2}_{E_0-\Delta E'/2}~dE_{2}dE_{3}~
        e^{i(E_1-E_{n+1})t}
        \\&
        \times
        \frac{(n-1)(n-2)}{2}
        Y(E_1,E_2)Y(E_2,E_3)Y(E_3,E_{n+1})
        D(E_1)\frac{\sin^2\Big(\pi e^{S_0}D(E_2)(E_2-E_3)\Big)}{\pi^2(E_2-E_3)^2}D(E_{n+1})
        \\&
        =\frac{(n-1)(n-2)}{2}\Big(\frac{\Delta E'}{4\pi E_0^{1/2}}\Big)^{n}
        \Big(\theta(t-t_0)-\theta(t-t_s)\Big)\frac{t^2-2tt_0+t_k^2}{T_H^2}.
        \end{split}
\end{equation}
The second term is
\begin{equation}
    \begin{split}
        &
        \mathcal{N}^{-1}
        e^{S_0}
        \Big(\frac{\Delta E'}{4\pi E_0^{1/2}}        
        \Big)^{n-1}
        \int^{E_0+\Delta E/2}_{E_0-\Delta E/2} dE_1dE_{n+1}~
        e^{i(E_1-E_{n+1})t}
        2(n-1)Y(E_1,E_{n+1})
        D(E_{n+1})\frac{t_s-t_0}{\pi}
        \\&
        -\mathcal{N}^{-1}
        e^{S_0}
        \Big(\frac{\Delta E'}{4\pi E_0^{1/2}}        
        \Big)^{n-2}
        \int^{E_0+\Delta E/2}_{E_0-\Delta E/2} dE_1dE_{n+1}\int^{E_0+\Delta E'/2}_{E_0-\Delta E'/2}~dE_{2}~
        e^{i(E_1-E_{n+1})t}
        \\&
        \times
        (n-1)Y(E_1,E_2)Y(E_2,E_{n+1})
        \frac{\sin^2\Big(\pi e^{S_0}D(E_2)(E_1-E_2)\Big)}{\pi^2(E_1-E_2)^2}D(E_{n+1})
        \\&
        -\mathcal{N}^{-1}
        e^{S_0}
        \Big(\frac{\Delta E'}{4\pi E_0^{1/2}}        
        \Big)^{n-2}
        \int^{E_0+\Delta E/2}_{E_0-\Delta E/2} dE_1dE_{n+1}\int^{E_0+\Delta E'/2}_{E_0-\Delta E'/2}~dE_{2}~
        e^{i(E_1-E_{n+1})t}
        \\&
        \times
        (n-1)Y(E_1,E_2)Y(E_2,E_{n+1})
        D(E_1)\frac{\sin^2\Big(\pi e^{S_0}D(E_2)(E_2-E_{n+1})\Big)}{\pi^2(E_2-E_{n+1})^2}
        \\&
        =
        2(n-1)\Big(\frac{\Delta E'}{4\pi E_0^{1/2}}\Big)^{n}
        \Big(\theta(t-t_0)-\theta(t-t_s)\Big)\frac{t^2-2tt_0+t_s^2}{T_H^2}.
        \end{split}
\end{equation}
The third term is
\begin{equation}
    \begin{split}
    &
    \Big(\frac{\Delta E'}{4\pi E_0^{1/2}}\Big)^{n}
    \Big[
    2\int_{t_0}^{t_s}dt_l~\Big(\frac{\Delta E'}{2\pi}\Big)
    \frac{\langle \text{TFD} (t)|t_l\rangle\langle t_l|\text{TFD}(t)\rangle}{\langle     t_l|t_l\rangle}
    -\Big(\theta(t_s-t)-\theta(t-t_0)\Big)\Big]
    \\&
    =
    \Big(\frac{\Delta E'}{4\pi E_0^{1/2}}\Big)^{n}
    \int_{t_0}^{t_s}dt_l~
    \frac{1}{T_H^2}
    \Big[\text{Min}\Big[T_H,~|t_l- t|\Big]+\text{Min}\Big[T_H,~|t_l+t|\Big]\Big].
    \end{split}
\end{equation}
More explicitly, it is equal to
\begin{equation}
    \begin{split}
        &
        \Big(\frac{\Delta E'}{4\pi E_0^{1/2}}\Big)^{n}
        \frac{1}{T_H^2}
        \Big(2T_H(t_s-t_0)+\theta(t-t_0)(t-t_0)^2-(t+t_0)^2
        \\&
        -\theta(t-t_0-T_H)(t-t_0-T_H)^2/2+\theta(t+t_0-T_H)(t+t_0-T_H)^2/2-\theta(t-t_0+T_H)(t-t_0+T_H)^2/2
        \\&
        +(t+t_0+T_H)^2/2
        -\theta(t-t_s)(t-t_s)^2   
        +(t+t_s)^2
        +\theta(t-t_s-T_H)(t-t_s-T_H)^2/2
        \\&
        +\theta(t-t_s+T_H)(t-t_s+T_H)^2/2
        -\theta(t+t_s-T_H)(t+t_s-T_H)^2/2
        -(t+t_s+T_H)^2/2
        \Big).
        \end{split}
\end{equation}
Summing over these contribution, for $t>0$ and $T_H/2>t_s>t_0>0$, we have
\begin{equation}
    \begin{split}
        Q^L(n,t_0,t_s)^{(1)}
        &
        =
        \Big(\frac{\Delta E'}{4\pi E_0^{1/2}}\Big)^{n}\times
        \Big[\frac{(n-1)(n+2)}{2}
        \Big(\theta(t-t_0)-\theta(t-t_s)\Big)\frac{t^2-2t_0t+t_s^2}{T_H^2}     
        \\&
        +
        \int_{t_0}^{t_s}dt_l~
        \frac{1}{T_H^2}
        \Big(\text{Min}\Big[T_H,~|t_l- t|\Big]+\text{Min}\Big[T_H,~|t_l+t|\Big]\Big)
        \Big].     
    \end{split}
\end{equation}
For $O^L$, we have
\begin{equation}
    \begin{split}
        O^L(n,t_k)^{(1)}
        &=
        \Big(\frac{\Delta E'}{4\pi E_0^{1/2}}\Big)^{n}
        \frac{n(n-1)}{2}\Delta E' \frac{2}{3\pi}\frac{(t_s-t_0)^2(2t_s+t_0)}{T_H^2}.
    \end{split}
\end{equation}

%%%%%%%%%%%%%%%%%%%%%%%%%%%%%%%%%%%%%%%%%%%%%%%%%%%%
%%%%%%%%%%%%%%%%%%%%%%%%%%%%%%%%%%%%%%%%%%%%%%%%%%%%
%%%%%%%%%%%%%%%%%%%%%%%%%%%%%%%%%%%%%%%%%%%%%%%%%%%%
\subsection{Timeshift}\label{appendix:timeshift}

We present detailed calculations of finite $t_s/T_H$ correction to the $Q^\delta$ and $O^\delta$. The correction is
\begin{equation}
    \begin{split}
        Q^\delta(n,t,t_-,t+)^{(1)}
        &=
        \mathcal{N}^{-1}e^{4S_0}
        \Big(\Delta E'        
        \Big)^{n-3}
        \int^{E_0+\Delta E/2}_{E_0-\Delta E/2} dE_1dE_{n+1}\int^{E_0+\Delta E'/2}_{E_0-\Delta E'/2}~dE_{2}dE_{3}~
        e^{i(E_1-E_{n+1})t}
        \\&
        \times
        \frac{(n-1)(n-2)}{2}
        Y(E_1,E_2)Y(E_2,E_3)Y(E_3,E_{n+1})
        D(E_1)\langle D(E_2)D(E_3)\rangle_{\text{c}}D(E_{n+1})    
        \\&
        +
        \mathcal{N}^{-1}
        e^{3S_0}
        \Big(\Delta E'        
        \Big)^{n-2}
        \int^{E_0+\Delta E/2}_{E_0-\Delta E/2} dE_1dE_{n+1}\int^{E_0+\Delta E'/2}_{E_0-\Delta E'/2}~dE_{2}~
        e^{i(E_1-E_{n+1})t}
        \\&
        \times
        (n-1)Y(E_1,E_2)Y(E_2,E_{n+1})
        \Big(\langle D(E_1)D(E_2)\rangle_{\text{c}}D(E_{n+1})+D(E_1)\langle D(E_2)D(E_{n+1})\rangle_{\text{c}}
        \Big) 
        \\&
        +
        \mathcal{N}^{-1}
        e^{2S_0}
        \Big(\Delta E'       
        \Big)^{n-1}
        \int^{E_0+\Delta E/2}_{E_0-\Delta E/2} dE_1dE_{n+1}~
        e^{i(E_1-E_{n+1})t}
        Y(E_1,E_{n+1})
        \langle D(E_1)D(E_{n+1})\rangle_{\text{c}}.
    \end{split}
\end{equation}
The first term is
\begin{equation}
    \begin{split}
        &\mathcal{N}^{-1}e^{S_0}
        \Big(\Delta E'        
        \Big)^{n-1}
        \int^{E_0+\Delta E/2}_{E_0-\Delta E/2} dE_1dE_{n+1}~
        e^{i(E_1-E_{n+1})t}
        \frac{(n-1)(n-2)}{2}
        Y(E_1,E_{n+1})
        D(E_{n+1})\frac{t_-+t_+}{2\pi}
        \\&-
        \mathcal{N}^{-1}e^{2S_0}
        \Big(\Delta E'        
        \Big)^{n-3}
        \int^{E_0+\Delta E/2}_{E_0-\Delta E/2} dE_1dE_{n+1}\int^{E_0+\Delta E'/2}_{E_0-\Delta E'/2}~dE_{2}dE_{3}~
        e^{i(E_1-E_{n+1})t}
        \\&
        \times
        \frac{(n-1)(n-2)}{2}
        Y(E_1,E_2)Y(E_2,E_3)Y(E_3,E_{n+1})
        D(E_1)\frac{\sin^2\Big(\pi e^{S_0}D(E_2)(E_2-E_3)\Big)}{\pi^2(E_2-E_3)^2}D(E_{n+1})
        \\&
        =\frac{(n-1)(n-2)}{2}\Big(\Delta E'\Big)^{n}
        \Big(\theta(t+t_-)-\theta(t-t_+)\Big)        
        \Big(t^2+t(t_--t_+)+\frac{t_-^2+t_+^2}{2}
        \Big)\frac{1}{T_H^2}.
        \end{split}
\end{equation}
The second term is
\begin{equation}
    \begin{split}
        &
        \mathcal{N}^{-1}
        e^{S_0}
        \Big(\Delta E'        
        \Big)^{n-1}
        \int^{E_0+\Delta E/2}_{E_0-\Delta E/2} dE_1dE_{n+1}~
        e^{i(E_1-E_{n+1})t}
        2(n-1)Y(E_1,E_{n+1})
        D(E_{n+1})\frac{t_-+t_+}{2\pi}
        \\&
        -\mathcal{N}^{-1}
        e^{S_0}
        \Big(\Delta E'       
        \Big)^{n-2}
        \int^{E_0+\Delta E/2}_{E_0-\Delta E/2} dE_1dE_{n+1}\int^{E_0+\Delta E'/2}_{E_0-\Delta E'/2}~dE_{2}~
        e^{i(E_1-E_{n+1})t}
        \\&
        \times
        (n-1)Y(E_1,E_2)Y(E_2,E_{n+1})
        \frac{\sin^2\Big(\pi e^{S_0}D(E_2)(E_1-E_2)\Big)}{\pi^2(E_1-E_2)^2}D(E_{n+1})
        \\&
        -\mathcal{N}^{-1}
        e^{S_0}
        \Big(\Delta E'       
        \Big)^{n-2}
        \int^{E_0+\Delta E/2}_{E_0-\Delta E/2} dE_1dE_{n+1}\int^{E_0+\Delta E'/2}_{E_0-\Delta E'/2}~dE_{2}~
        e^{i(E_1-E_{n+1})t}
        \\&
        \times
        (n-1)Y(E_1,E_2)Y(E_2,E_{n+1})
        D(E_1)\frac{\sin^2\Big(\pi e^{S_0}D(E_2)(E_2-E_{n+1})\Big)}{\pi^2(E_2-E_{n+1})^2}
        \\&
        =
        2(n-1)\Big(\Delta E'\Big)^{n}
        \Big(\theta(t+t_-)-\theta(t-t_+)\Big)
        \Big(t^2+t(t_--t_+)+\frac{t_-^2+t_+^2}{2}\Big)
        \frac{1}{T_H^2}.
        \end{split}
\end{equation}
The third term is
\begin{equation}
    \begin{split}
    &
    \Big(\Delta E'\Big)^{n}
    \Big[
    \int_{-t_-}^{t_+}dt_l~\Big(\frac{\Delta E'}{2\pi}\Big)
    \frac{\langle \text{TFD} (t)|\delta\rangle\langle \delta|\text{TFD}(t)\rangle}{\langle     \delta|\delta\rangle}
    -\Big(\theta(t+t_-)-\theta(t-t_+)\Big)\Big]
    \\&
    =
    \Big(\Delta E'\Big)^{n}
    \int_{-t_-}^{t_+}d\delta~
    \frac{1}{T_H^2}
    \text{Min}\Big[T_H,~|\delta- t|\Big].
    \end{split}
\end{equation}
More explicitly,
\begin{equation}
    \begin{split}
        &
        \Big(\Delta E'\Big)^{n}
        \Big(\frac{t_-+t_+}{T_H}+\frac{1}{T_H^2}
        \Big(\theta(t+t_-)(t+t_-)^2-\theta(t-T_H+t_-)(t-T_H+t_-)^2/2
        \\&
        -\theta(t+T_H+t_-)(t+T_H+t_-)^2/2-\theta(t-t_+)(t-t_+)^2+\theta(t-T_H-t_+)(t-T_H-t_+)^2/2
        \\&
        +\theta(t+T_H-t_+)(t+T_H-t_+)^2/2\Big).        
        \end{split}
\end{equation}

%%%%%%%%%%%%%%%%%%%%%%%%%%%%%%%%%%%%%%%%%%%%%%%%%%%%
%%%%%%%%%%%%%%%%%%%%%%%%%%%%%%%%%%%%%%%%%%%%%%%%%%%%
%%%%%%%%%%%%%%%%%%%%%%%%%%%%%%%%%%%%%%%%%%%%%%%%%%%%

\subsection{Replica Calculations}\label{appendix:replica}
%%%%%%%%%%%%%%%%%%%%%%%%%%%%%%%%%%%%%%%%%%%%%%%%%%%%
%%%%%%%%%%%%%%%%%%%%%%%%%%%%%%%%%%%%%%%%%%%%%%%%%%%%
%%%%%%%%%%%%%%%%%%%%%%%%%%%%%%%%%%%%%%%%%%%%%%%%%%%%

We write the short and long-interval limit results for general $n$.

\subsubsection*{Short Interval Limit}
When $t_0,~t_s,~t_-,~t_+\ll T_H$, we have
\begin{equation}
    \begin{split}
        Q^L(n,t_0,t_s)^{(0)}
        &=
        \Big(\frac{\Delta E'}{4\pi E_0^{1/2}}        
        \Big)^n\Big(\theta(t+t_s)-\theta(t+t_0)-\theta(t-t_s)+\theta(t-t_0)\Big),
        \\
        O^L(n,t_0,t_s)^{(0)}
        &=
        \Big(\frac{\Delta E'}{4\pi E_0^{1/2}}        
        \Big)^n\frac{t_s-t_0}{\pi}\Delta E',
        \\
        Q^\delta(n,t_-,t_+)^{(0)}
        &=
        (\Delta E')^n\Big(\theta(t+t_-)-\theta(t-t_+)\Big),
        \\
        O^\delta(n,t_-,t_+)^{(0)}
        &=
        (\Delta E')^n\frac{t_-+t_+}{2\pi}\Delta E'.
    \end{split}
\end{equation}
%%%%%%%%%%%%%%%%%%%%%%%%%%%%%%%%%%%%%%%%%%%%%%%%%%%%
%%%%%%%%%%%%%%%%%%%%%%%%%%%%%%%%%%%%%%%%%%%%%%%%%%%%
%%%%%%%%%%%%%%%%%%%%%%%%%%%%%%%%%%%%%%%%%%%%%%%%%%%%
\subsubsection*{Long Interval Limit}
When $t_s-t_0,~t_-,~t_+\gg T_H$, we have
\begin{equation}
    \begin{split}
        Q^L(n,t_0,t_s)
        &=
        \Big(\frac{(t_s-t_0)\Delta E'}{4\pi^2 e^{S_0}D(E_0) E_0^{1/2}}        
        \Big)^n,
        \\
        O^L(n,t_0,t_s)
        &=
        \Big(\frac{(t_s-t_0)\Delta E'}{4\pi^2 e^{S_0}D(E_0) E_0^{1/2}}        
        \Big)^ne^{S_0}D(E_0)\Delta E',
        \\
        Q^\delta(n,t_-,t_+)
        &=
        \Big(\frac{(t_-+t_+)\Delta E'}{2\pi e^{S_0}D(E_0)}        
        \Big)^n,
        \\
        O^\delta(n,t_-,t_+)
        &=
         \Big(\frac{(t_-+t_+)\Delta E'}{2\pi e^{S_0}D(E_0)}        
        \Big)^ne^{S_0}D(E_0)\Delta E'.
    \end{split}
\end{equation}

\subsection{Black and White hole probability}

We write the black hole and white hole probability including the first correction from finite $(t_s-t_0)/T_H$ and $t_s/T_H$ effect. They are given by
\begin{equation}
    \begin{split}
    P^{\text{BH}}(t)^{(0)+(1)}=\int_0^{T_H/2}d\delta'~P^+(\delta')^{(0+1)}
    &
    =\theta(T_H/2-t)\Big(1-\frac{5t^2-T_H t+T_H^2/4}{2T_H^2}\Big)
    \\
    &+\frac{1}{8}+\frac{1}{T_H^2}
        \Big(t^2-tT_H/2+\theta(t-T_H)(t-T_H)^2/2
        \\&-\theta(t-T_H/2)(t-T_H/2)^2+\theta(t-3T_H/2)(t-3T_H/2)^2/2\Big),
    \\
    P^{\text{WH}}(t)^{(0)+(1)}=\int_{-T_H/2}^0d\delta'~P^-(\delta')^{(0+1)}
    &
    =\theta(T_H/2-t)\frac{3t^2-T_H t-T_H^2/4}{2T_H^2}
    \\&
    +\frac{1}{8}+\frac{1}{T_H^2}
        \Big(tT_H/2-\theta(t-T_H/2)(t-T_H/2)^2/2
        \\&
        +\theta(t-T_H)(t-T_H)^2/2\Big).
    \end{split}
\end{equation}

% \newpage

\bibliographystyle{jhep}
\bibliography{references}

\end{document}